# Time-Reversed Superfluorescence in a Polaronic Quantum Material

Arnab Ghosh[1], Patrick Brosseau[1], Dmitry N. Dirin[2,3], Maksym V. Kovalenko[2,3], and Patanjali Kambhampati[1]*

[1]Department of Chemistry, McGill University, Montreal, H3A 0B8, Canada

[2] Department of Chemistry and Applied Biosciences, ETH Zürich, Switzerland

[3] Laboratory for Thin Films and Photovoltaics, Empa - Swiss Federal Laboratories for Materials Science and Technology, Switzerland

*pat.kambhampati@mcgill.ca

**Abstract**

Superfluorescence—the cooperative burst of spontaneous emission from an ensemble of dipoles—arises when microscopic oscillators spontaneously synchronize their phases. Here we show that this process can be reversed in time within quantum materials. Coherent multidimensional spectroscopy of halide perovskite quantum dots reveals a delayed cooperative absorption burst—the mirror image of superfluorescent emission—driven by transient polaron fields that phase-lock unit-cell dipoles within 100 fs. The effect scales systematically with quantum-dot size and halide composition, reaching near-unity coherence fidelity even at 300 K. A microscopic exciton–polaron model reproduces the buildup and decay of the coherent state, identifying lattice polarons as the mediators of synchronization. These results demonstrate that many-body temporal coherence can self-organize and persist at room temperature, opening routes toward engineered collective optical states and superabsorbing quantum devices.

Light–matter interactions can drive ensembles of emitters into collective quantum states that transcend the behavior of individual dipoles. Two paradigmatic examples are Dicke superradiance (SR) [1-5] and superfluorescence (SF) [6, 7]. SR arises from static spatial coherence: dipoles pre-aligned by their common radiation field emit a burst whose intensity scales superlinearly with emitter number. SF, by contrast, embodies many-body temporal coherence: an inverted but randomly phased ensemble spontaneously synchronizes through dipole–dipole coupling mediated by the shared field, generating order from microscopic disorder and releasing a delayed cooperative burst. Together, these processes define the canonical landscape of collective quantum optics [5-11], with realizations most recently in perovskite quantum dots [3, 6, 11, 12].

Although emission and absorption are often portrayed as time-reversed processes, they are not strict microscopic counterparts [13]. Absorption arises from the interference of an induced polarization with the driving optical field, whereas spontaneous emission originates from population decay seeded by vacuum fluctuations. Nevertheless, the same dipole–dipole couplings that amplify cooperative emission can also enhance collective absorption, establishing a formal reciprocity between the two.

For superradiance (SR) [3, 8], this conjugate process—superabsorption (SA)—has been realized across several platforms[10, 14-17] including most recently perovskite quantum dots[11, 12], confirming SR and SA as reciprocal cooperative responses. For superfluorescence (SF) [6, 7], however, the situation is fundamentally different. Because SF involves the spontaneous buildup of coherence rather than preexisting phase alignment, its conjugate process would not appear as prompt enhanced absorption, but as a delayed cooperative absorption burst—a phenomenon long anticipated in theory [1] yet never observed in any medium [11].

Here we demonstrate the long-sought conjugate process to superfluorescence. Using coherent multidimensional spectroscopy, we directly capture the ultrafast cooperative absorption burst that mirrors superfluorescent emission in ensembles of isolated perovskite quantum dots. Rather than a prompt single-particle response, we observe a delayed collective uptake of energy, gated by transient polaron fields that synchronize unit-cell dipoles into a coherently phased array. The buildup unfolds within 50–100 femtoseconds, displays systematic size and composition dependence across $CsPbBr_3$ and $CsPbI_3$, and reaches fidelities approaching 99% at 300 K. A microscopic exciton–polaron model quantitatively reproduces the formation and dephasing of the coherent state, identifying lattice polarons as the enabling quasiparticles that mediate dipole synchronization. These results constitute the first real-time observation of the cooperative absorption conjugate to superfluorescence, revealing a lattice-driven pathway to ultrafast many-body coherence and opening routes toward superabsorbing quantum devices, ultrafast memories, and coherence-engineered photonic states.

**Figure 1a** schematically depicts the Dicke model of collective absorption and emission, visualized in terms of microscopic and macroscopic dipoles. Each arrow represents the transition dipole moment of an individual emitter. In the dilute regime ($D > D_c$), where the inter-emitter spacing $D$exceeds the critical coupling distance $D_c$, the dipoles oscillate with random phases and their individual fields add incoherently, producing only weak, independent emission. The total macroscopic dipole moment—shown as the vector sum of the individual dipoles—is therefore small. In the cooperative regime ($D < D_c$), the emitters lie within one another's near-field or radiation-coupling range, allowing the shared electromagnetic field to synchronize their phases. As the individual dipoles align, the ensemble develops a single, large macroscopic dipole whose amplitude scales as $N$, producing emission or absorption intensities proportional to $N^2$. This transition from randomly phased oscillators to a coherently phased array marks the onset of a Dicke state, the physical basis of both superradiance and superabsorption.

**Figure 1b** extends this conceptual framework to the time-reversed superfluorescent (TR-SF) process that forms the core of this study. Here, the inter-dipole spacing remains fixed—well within the regime where radiative and lattice couplings are significant—but the ensemble initially consists of randomly phased dipoles ($|\ 0\rangle \rightarrow |\ X_1\rangle$) following excitation by a femtosecond optical pulse. During the ensuing delay time (Δt), the optical perturbation and the transient polaronic lattice polarization act together as a coupling field, progressively synchronizing the oscillating dipoles without changing their spatial arrangement. This temporal self-organization converts microscopic phase disorder into macroscopic order: as the dipoles align in phase, their individual moments sum coherently to form a large macroscopic dipole, and the ensemble collectively absorbs energy in a delayed burst. The process thus represents the absorption-channel conjugate of superfluorescent emission—not a simple mirror of single-particle absorption, but a time-dependent cooperative absorption event arising from many-body phase alignment mediated by transient lattice fields.

The material platform is $CsPbBr_3$, a lead–halide perovskite composed of corner-sharing $PbX_6$ octahedra (X = Br) that form a flexible inorganic framework (**Fig. 1c**) [18-20]. At room temperature, $CsPbBr_3$ adopts a distorted orthorhombic structure with dynamic octahedral tilts and halide displacements. This soft, polar lattice exhibits pronounced anharmonicity and symmetry fluctuations. Such liquid-like, dynamically disordered behavior enhances Fröhlich-type exciton–lattice coupling, promoting polaron formation on femtosecond timescales [21-32].

**Figure 1b** extends this framework to the time-reversed superfluorescent (TR-SF) process that forms the core of this study. Here, the spatial arrangement of dipoles remains fixed within a two-dimensional quantum lattice, but the ensemble begins as a sea of randomly phased oscillators ($|0\rangle \rightarrow |X_1\rangle$) driven by an ultrafast optical perturbation. During the ensuing delay period (Δt), correlated lattice and electronic motions spontaneously organize the system: the transient polaronic polarization field that forms through exciton–lattice coupling feeds back onto the electronic dipoles, aligning their phases across the array.

In this regime, the lattice ceases to be a passive dephasing bath and becomes an active mediator of coherence, dynamically enhancing the effective dipole–dipole coupling. The ensemble evolves from microscopic phase disorder to macroscopic quantum order, assembling a coherently phased array whose collective dipole moment grows in time and produces a delayed cooperative absorption burst. This process realizes the absorption-channel conjugate of superfluorescence, an emergent many-body phenomenon in which lattice-driven synchronization and electronic coherence coalesce to generate a macroscopic quantum oscillator from a nominally disordered solid.

**Figure 1d** compares the vibrational spectral signatures that govern the cooperative behavior of perovskite and CdSe quantum dots. In perovskite quantum dots, the broadband quasi-elastic continuum dominating the low-frequency Raman spectra reflects a dense, anharmonic spectral density arising not from discrete phonon modes but from collective liquid-like polaron dynamics—the correlated motion of charge carriers and lattice distortions. The low-energy Raman spectrum of perovskite is zero-frequency–dominated with a Debye-type roll-off, well captured by an overdamped Drude–Debye spectral density $J(\omega) = 2\lambda\,\omega\gamma/(\omega^2 +$ $\gamma^2$)supplemented by a weak sub-Ohmic tail with a soft cutoff near 30–40 meV, consistent with relaxational polaronic lattice dynamics rather than discrete phonons. In stark contrast, CdSe quantum dots display only narrow, well-resolved phonon lines on a weak background, characteristic of a rigid covalent lattice with negligible anharmonic coupling.

The low-frequency continuum in perovskites thus signifies a regime where lattice and electronic degrees of freedom are dynamically entangled, giving rise to correlated motion that mediates emergent many-body coherence. In this picture, the lattice is not a passive thermal bath but an active quantum medium whose reorganization generates polarization fields that transiently confine and phase-couple excitons across the lattice. Such feedback between the lattice polarization and electronic coherence establishes a self-consistent polaron field—a collective mode that bridges microscopic exciton–phonon interactions and macroscopic dipole synchronization. This coupling framework moves beyond conventional dephasing or renormalization, positioning the perovskite lattice as a driven, self-organizing system capable of amplifying coherent order. In contrast, CdSe represents the limiting case of a rigid, weakly polar semiconductor, where phonons merely scatter excitons without mediating collective order. The juxtaposition delineates two regimes of light–matter interaction: the phonon-limited coherence of covalent lattices versus the polaron-enabled cooperativity of soft, polar lattices—providing the essential microscopic context for the emergence of time-reversed superfluorescence as a lattice-driven, many-body quantum phenomenon.

**Figure 2(a)** schematically illustrates the coherent multidimensional spectroscopy (CMDS) [13, 33, 34] pulse sequence used to directly observe time-reversed superfluorescence. In this experiment, three phase-locked femtosecond fields $E_1$, $E_2$, and $E_3$interact with the quantum-dot ensemble to generate a measurable signal field through a third-order nonlinear polarization, $P^{(3)}(t)$. The first field, $E_1$, induces a linear polarization $P^{(1)}(t_1)$corresponding to the ground–exciton coherence $|\,0\rangle \leftrightarrow |\,X_1\rangle$. The second field, $E_2$, converts this coherence into an excited-state population in $|\,X_1\rangle$that evolves during the population period $t_2$. During this interval, the dipoles interact through the lattice field, and their mutual correlations—whether incoherent or cooperative—develop in real time. The third field, $E_3$, then drives the population back into a radiating coherence that generates the nonlinear polarization $P^{(3)}(t_3)$. The emitted signal field, $E_{\mathrm{sig}} \propto P^{(3)}$, is heterodyne-detected with a phase-stable local oscillator to recover both its amplitude and phase. By scanning the delays $t_1$and $t_3$and Fourier-transforming along both excitation and detection axes, CMDS constructs a two-dimensional frequency map that correlates the linear absorption ($P^{(1)}$) with the nonlinear response ($P^{(3)}$). The experiments are done in the pump/probe geometry to produce absorptive spectra as described in the Supplementary Information (**SI**) and our prior work [23-25, 35, 36].

In conventional materials, the linear and nonlinear polarizations are uncorrelated beyond single-particle dephasing and relaxation, and the nonlinear absorption arises promptly within the instrument response. In contrast, in perovskite quantum dots, the evolution of the $t_2$population reveals the emergence of a delayed cooperative nonlinear response, marking the buildup of collective phase correlations—the real-time signature of time-reversed superfluorescence.

**Figure 2b** illustrates how perovskite quantum dots evolve into the time-reversed superfluorescent (TR-SF) regime during the population period $t_2$. After excitation by $E_1$and $E_2$, the ensemble initially consists of randomly phased dipoles within the excitonic manifold $|\,X_1\rangle$. As this population evolves, transient polaron fields arise through exciton–lattice coupling, generating correlated local distortions that serve as a mediating field linking dipoles across the quantum dot array. These polaron fields establish long-range interactions that progressively synchronize dipole phases, transforming an initially disordered ensemble into a coherently phased array. When the third field $E_3$subsequently interacts with this synchronized population, the induced dipole moment is strongly amplified, producing a third-order polarization $P^{(3)}(t_3)$ far greater than that of independent absorbers. The resulting signal field, proportional to $P^{(3)}$, manifests as a delayed cooperative absorption burst—the time-reversed analog of superfluorescent emission.

This process represents the spontaneous buildup of many-body temporal coherence: during $t_2$, dipole–dipole correlations mediated by the lattice evolve from incoherence toward collective order, causing the nonlinear signal to grow in time rather than the expected step function response. The enlarged dipole moment at the $E_3$ interaction and the amplified $P^{(3)}$ emission visualize the emergence of cooperative absorption, revealing that the ensemble has undergone a self-organized transition from randomly phased dipoles to a single macroscopic quantum oscillator—the defining hallmark of time-reversed superfluorescence.

The absorptive CMDS spectra of 18 nm $CsPbBr_3$ quantum dots reveal this effect unambiguously (**Fig. 2c**). At $t_2 = 20$fs—immediately following the 10 fs excitation pulse—the CMDS map displays a single absorptive peak corresponding to the band-edge exciton $|X_1\rangle$. The signal is normalized at this early delay, capturing the ensemble directly after excitation but before any coherent evolution begins. Excitation is resonant with $|X_1\rangle$to avoid hot-exciton relaxation [35] and isolate the intrinsic, polaron-mediated dynamics.

By $t_2 = 100\text{fs}$, the CMDS amplitude increases to roughly 1.5 × its initial value, accompanied by a distinct distortion of the peak shape. This delayed amplification marks the cooperative absorption burst characteristic of time-reversed superfluorescence. The evolving lineshape reflects liquid-like spectral diffusion driven by lattice reorganization in halide perovskites [23-25]. By $t_2 = 300\text{fs}$, the signal amplitude relaxes toward its initial level, completing a coherence cycle in which the ensemble evolves from independent absorbers to a collectively phased state and back. Subtle asymmetries in the long-delay lineshape trace the finite lifetime of the polaron field and its imprint on excitonic lineshape dynamics.

Control experiments using CdSe quantum dots emphasize the material specificity of the phenomenon (**Fig. 2d**). CdSe, whose rigid covalent lattice supports only discrete optical phonons [37], exhibits no delayed cooperative feature—only prompt, uncorrelated absorption. The contrast does not simply reflect a difference in coupling strength but in the nature of the coupled degrees of freedom. In CdSe, electrons and phonons remain largely separable: the lattice provides an independent bath of quantized vibrations that modulate but do not reorganize the electronic polarization. In perovskite quantum dots, by contrast, the electronic and lattice coordinates evolve correlatively, forming transient polaron quasiparticles whose collective fields feed back on the dipoles themselves. This correlated motion creates a self-consistent polarization landscape that can synchronize dipoles in time—a dynamical, many-body response absent in conventional covalent semiconductors.

Together, the panels of **Fig. 2** trace, in real time, the transformation of unit-cell dipoles from a disordered ensemble of independent absorbers into a coherently phased array governed by collective quantum dynamics. The CMDS sequence directly visualizes how the lattice field, initially acting as a fluctuating background, reorganizes into a correlated polaronic mode that couples dipoles across the ensemble, converting microscopic incoherence into macroscopic order. The delayed growth and subsequent decay of the nonlinear signal thus capture the full life cycle of a collective quantum state—its birth through lattice-mediated synchronization, its peak cooperative absorption, and its relaxation through dephasing of the polaron field. These dynamics constitute the first direct, time-resolved observation of a lattice-gated, time-reversed superfluorescent process in any solid-state material, establishing a new experimental paradigm for probing emergent many-body coherence in the time domain.

**Figure 3** quantifies the evolution of time-reversed superfluorescence (TR-SF) as a function of perovskite quantum-dot size and composition, revealing how lattice softness, quantum confinement, and exciton–phonon coupling collectively determine the buildup and fidelity of cooperative absorption. In **Fig. 3a**, the integrated peak volume CMDS dynamics of the band-edge exciton ($|X_1\rangle$) provide a direct temporal view of the cooperative absorption process. Each trace shows the normalized third-order signal amplitude as a function of population delay ($t_2$) for $CsPbBr_3$ quantum dots spanning edge lengths from 4.9 to 18.4 nm, together with a representative 15 nm $CsPbI_3$ sample. The instrument response function (IRF), shown in gray, defines the 20 fs temporal resolution of the measurement, while the full pulse spectrogram is presented in the **SI**.

All perovskite samples exhibit a delayed rise in signal intensity, peaking between 80 and 140 fs after excitation—signifying the ultrafast onset of collective absorption mediated by polaron formation. The rise occurs more rapidly and reaches higher amplitude in smaller $CsPbBr_3$ quantum dots, where stronger quantum confinement enhances exciton–phonon coupling and accelerates lattice reorganization. In contrast, larger dots display slower but more gradual dynamics, reflecting the rephasing of dipoles across an expanded coherence volume. $CsPbI_3$ shows a slightly slower buildup yet achieves a higher steady-state amplitude, consistent with its softer, more polarizable lattice that sustains stronger cooperative coupling.

The black trace in **Fig. 3a** shows the CdSe control, which exhibits an IRF-limited step-function response—prompt, featureless, and entirely non-cooperative. This absence of delayed dynamics does not arise from weak coupling per se but from its fundamentally different character: in CdSe, the electrons and lattice vibrations remain separable degrees of freedom, producing conventional phonon-limited absorption without collective feedback. By contrast, in perovskite quantum dots, electron–lattice interactions generate correlated polaron fields that reorganize the lattice polarization and feed back on the dipole ensemble, enabling the cooperative absorption burst. The comparison thus isolates the essential physics of time-reversed superfluorescence as a correlated many-body response, not merely a stronger form of single-particle coupling. Collectively, these dynamics demonstrate that the kinetics of TR-SF scale systematically with quantum-dot size and halide composition through their modulation of lattice polarizability, polaron formation dynamics, and the ensuing development of correlated many-body coherence.

**Figure 3b** quantifies how the cooperative absorption kinetics evolve across perovskite quantum-dot size and composition. The extracted rise-time rate constants, $k_r = 1/\tau_r$, measure how rapidly the time-reversed superfluorescence (TR-SF) signal builds as dipoles become phase-locked through the transient polaron field. Each data point is obtained from an exponential fit to the rising edge of the integrated CMDS peak amplitude in **Fig. 3a.** For $CsPbBr_3$ quantum dots, $k_r$ decreases systematically from approximately $1.4 \times 10^{13}$ $s^{-1}$ for 5 nm dots to $7 \times 10^{12}$ $s^{-1}$ for 18 nm dots, demonstrating that the rate constant for cooperative absorption slows with increasing dot size. This inverse scaling reflects the expanding spatial domain over which the transient polaron field must synchronize dipole phases: smaller dots reorganize their lattice collectively within a single vibrational period, whereas larger dots require longer reconfiguration times for the polaron field to coherently couple all unit-cell dipoles across their volume.

The overall scaling behavior, $k_r \propto L^{-1.0\pm0.1}$, captures the intrinsic polaron-field formation timescale that governs the cooperative absorption burst. The single $CsPbI_3$ data point lies slightly below the $CsPbBr_3$ series, consistent with its softer lattice, lower phonon frequency, and higher dielectric polarizability—all of which reduce the restoring force for exciton–lattice coupling and thus yield a smaller $k_r$. The solid lines represent fits to the microscopic exciton–polaron model, which quantitatively reproduce both the magnitude and size dependence of $k_r$. Collectively, these results establish that the cooperative absorption dynamics in halide perovskite quantum dots are governed not by extrinsic relaxation or trapping but by the intrinsic timescale of lattice-mediated dipole synchronization—a direct dynamical manifestation of correlated exciton–polaron many-body coherence.

**Figure 3c** quantifies the cooperative enhancement in absorption through the peak volume of the band-edge exciton signal, $A_{\mathrm{max}}$, extracted from the tail normalized CMDS dynamics. For $CsPbBr_3$ quantum dots, $A_{\mathrm{max}}$systematically increases with decreasing size, rising from approximately 1.35 for 25 nm dots to nearly 1.70 for 5 nm dots when normalized to the early-time ($t_2 = 20$ fs) response. This monotonic trend reflects that smaller quantum dots sustain stronger local electric fields and larger exciton–phonon coupling constants, enabling faster polaron formation and tighter phase synchronization among dipoles. In the cooperative framework, the amplitude enhancement scales with the effective number of synchronized dipoles, $A_{\max} \propto N_{\text{eff}}^{1/2}$; the observed variation therefore corresponds to an approximate twofold increase in $N_{\text{eff}}$across the studied size range.

The $CsPbI_3$ data point lies above the $CsPbBr_3$ series, consistent with its higher lattice polarizability and stronger Fröhlich coupling—both of which enhance collective phase alignment and produce a more pronounced cooperative response. The black CdSe control, shown for comparison, remains constant at unity, confirming purely independent absorption with no collective buildup. The solid line through the perovskite data represents a fit to the scaling relation $A_{\max} \propto L^{-0.3\pm0.1}$, in quantitative agreement with a microscopic exciton–polaron model that links the cooperative amplitude to the strength of exciton–lattice coupling and the spatial extent of phase synchronization.

**Figure 3d** presents the simulated fidelity of phase synchronization, $R_{\infty}$, derived from the microscopic exciton–polaron Hamiltonian for each experimentally studied quantum-dot size and composition. $R_{\infty}$quantifies the degree of long-range phase alignment that the ensemble attains after cooperative absorption—the fraction of dipoles that remain mutually phase-locked once the transient polaron field has fully developed. In the model, each dipole experiences both a local diffusive environment and a global coupling mediated by the lattice polarization; as the coupling dominates, their individual oscillations converge toward a common phase, producing a single macroscopic dipole. $R_{\infty} = 1$corresponds to perfect synchronization, while smaller values reflect partial coherence limited by dephasing and lattice disorder.

$$R(t) = \frac{1}{N}\left|\sum_{j=1}^{N} e^{i\Phi_j(t)}\right|$$

which quantifies the degree of phase alignment among all dipoles in the ensemble, with $R_\infty = 1$ corresponding to perfect synchronization. The simulations reproduce both the experimental dynamics and amplitude trends across the full size range. For $CsPbBr_3$ quantum dots, $R_\infty$ increases monotonically from $0.92 \pm 0.03$ at 12 nm to $0.95$ at 15 nm, $0.96$ at 18 nm, $0.97$ at 22 nm, and $0.98$ at 25 nm, reflecting the progressive stabilization of coherent phase locking as the exciton–lattice correlation volume expands.

The 15 nm $CsPbI_3$ ensemble achieves $R_\infty = 0.99 \pm 0.01$even at 300 K, demonstrating nearly perfect collective alignment under ambient conditions. This exceptional fidelity arises from the iodide lattice's high dielectric polarizability and dense low-frequency phonon spectrum, which together strengthen the lattice-mediated coupling that locks dipole phases. Across the $CsPbBr_3$ size series, $R_\infty$increases steadily with dot size, indicating that larger coherence volumes support more stable synchronization. In contrast, simulations based on CdSe parameters yield $R_\infty < 0.2$, consistent with the experimentally observed prompt, uncorrelated absorption.

**Figure 3e** reveals how the synchronized state in perovskite quantum dots dephases following the cooperative absorption burst. The experimental FFT spectra are obtained by Fourier transforming the decay portion of the CMDS kinetics (100–400 fs) after the TR-SF maximum. These spectra represent the energy distribution of dephasing components associated with the relaxation of the coherent, time-reversed superfluorescent state. For all $CsPbBr_3$ samples, the FFT magnitude displays a smooth roll-off from 0 to ≈ 40 meV, reflecting a broad continuum of lattice-coupled dephasing channels rather than discrete vibrational resonances.

The spectral weight progressively shifts toward lower energies and narrows slightly with increasing quantum-dot size, consistent with reduced lattice disorder, weaker scattering, and longer dephasing times $T_2^*$. The $CsPbI_3$ sample exhibits an even steeper low-energy roll-off and narrower bandwidth, indicative of its softer lattice and slower polarization response, which together sustain coherence for the longest times. The monotonic evolution of the FFT envelope across the halide and size series demonstrates that the coherence decay is governed by phonon-mediated relaxation of the collective polaron field that originally synchronized the dipoles.

**Figure 3f** shows the theoretical FFT spectra computed from the microscopic exciton–polaron model using experimentally constrained phonon spectral densities. The simulations accurately reproduce both the 0–40 meV bandwidth and its systematic narrowing with increasing quantum-dot size and lattice softness. In this framework, the decay of the TR-SF signal arises from a continuum of collective relaxation modes of the coupled exciton–lattice system. The cutoff energy of this continuum reflects the balance between lattice inertia and the strength of dipole–dipole coupling mediated by the polaron field. The close agreement between experiment and theory confirms that the TR-SF decay originates from coherent relaxation of a collective polaron mode rather than independent single-particle scattering.

The measured and simulated FFT spectra together show that the TR-SF decay originates from the coherent relaxation of a collective polaron field—not from single-particle scattering. The slight mismatch between experiment and theory in $CsPbI_3$ reflects composition-specific anharmonicity and dielectric screening, not missing interactions, and leaves the overall scaling intact. The 0–40 meV spectral envelope thus maps the dephasing landscape of the synchronized state, directly linking lattice softness, polaronic coupling, and collective coherence in a single physical picture.

We have demonstrated the time reversal of superfluorescence in perovskite quantum dots, revealing that collective absorption can emerge from polaron-driven synchronization of dipoles. This establishes a new regime of cooperative light–matter interaction in solids, where lattice motion actively generates and sustains many-body coherence at room temperature. By linking coherence fidelity to lattice softness and exciton–polaron coupling, these results define a framework for designing quantum materials that self-organize into coherent states—opening routes toward superabsorbing devices, ultrafast memories, and coherence-based quantum energy architectures.

**Figures**.

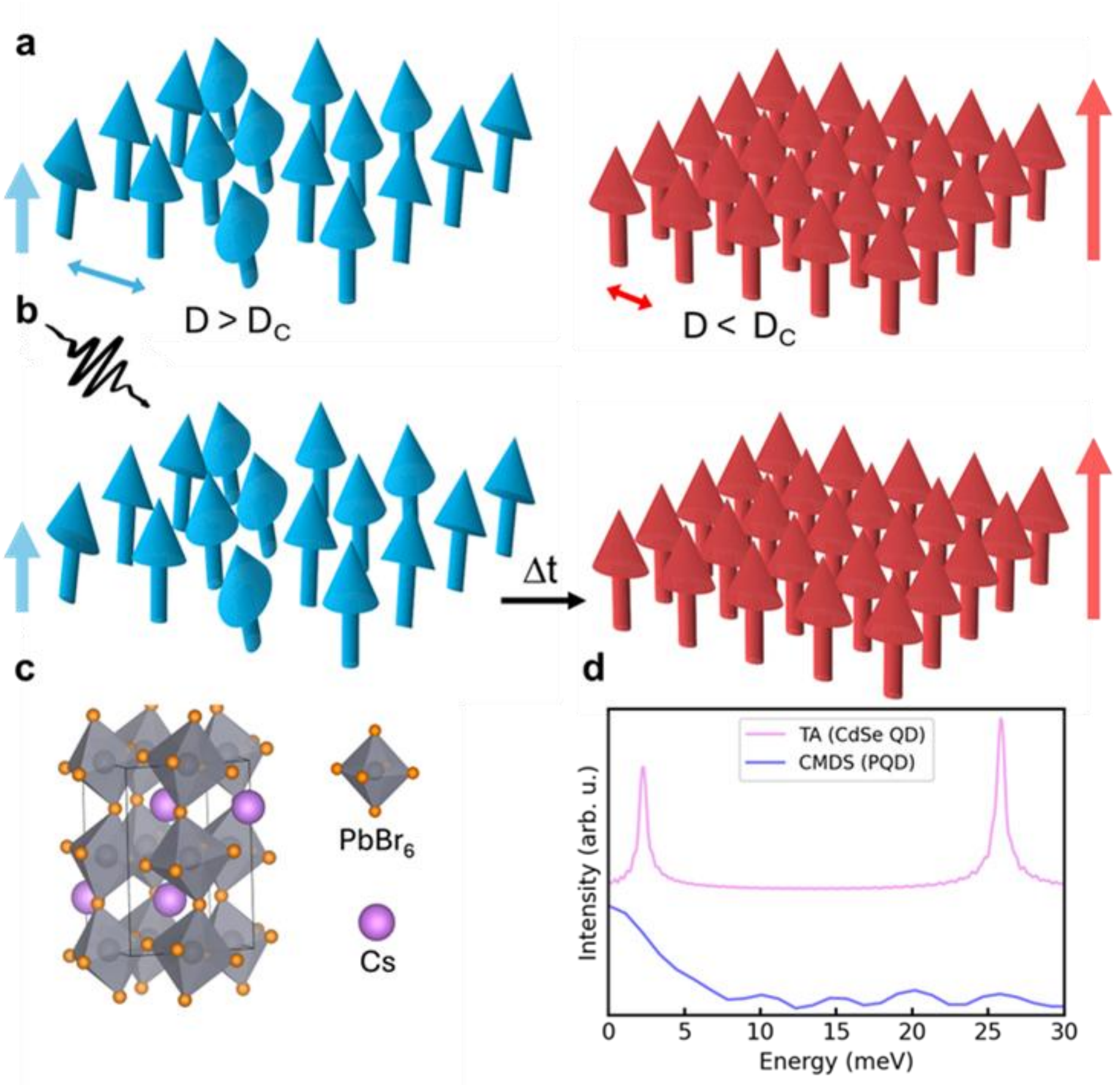

**Fig. 1 | Collective coherence in dipole ensembles.** a, For ensembles larger than the cooperative domain size ($D > D_c$), dipoles emit independently and remain randomly phased. When the coupling exceeds the critical threshold ($D < D_c$), interactions through the shared radiation field align dipoles, producing a coherent Dicke state that manifests as superradiance in emission or, by time reversal, cooperative absorption. b, In superfluorescence, an inverted but randomly phased ensemble spontaneously synchronizes through dipole–dipole coupling, yielding a delayed collective burst. Its time-reversed counterpart—long anticipated but never observed—would appear as a delayed burst of absorption. c, The soft, polar lattice of lead-halide perovskites reorganizes on femtosecond timescales, generating transient polaron fields that can synchronize dipoles into a coherently phased array. d, Raman-constrained spectral densities reveal this liquid-like lattice response in perovskite quantum dots, absent in covalent CdSe, providing the dynamical environment that enables cooperative quantum states in the solid phase.

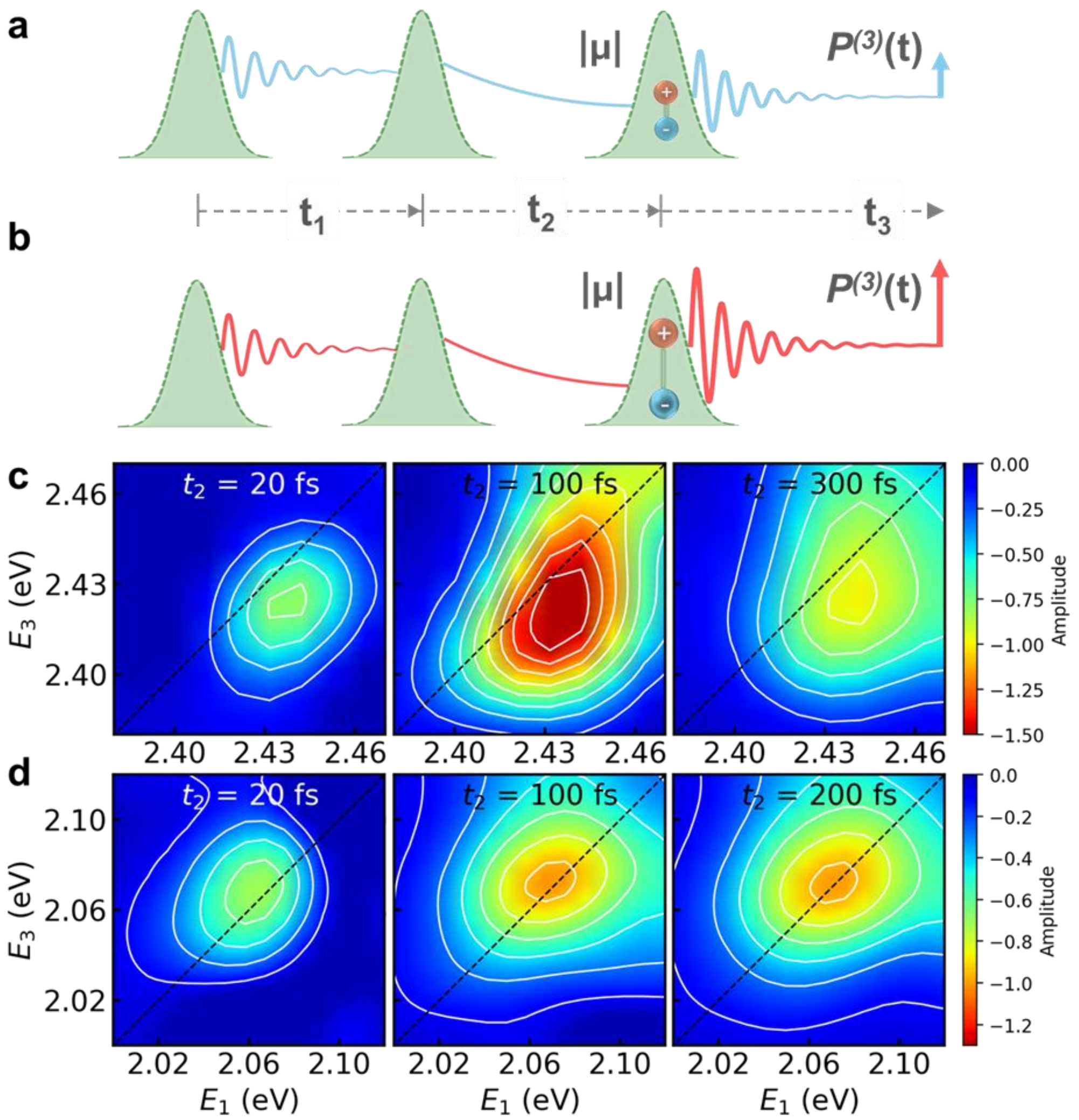

**Fig. 2 | Coherent multidimensional spectroscopy reveals time-reversed superfluorescence.** a, In the CMDS pulse sequence, three phase-coherent fields $E_1$, $E_2$, and $E_3$generate a third-order nonlinear polarization $P^{(3)}$, whose evolution with population time $t_2$maps dipole–dipole correlations. b, Following excitation by $E_1$and $E_2$, initially incoherent dipoles evolve under transient polaron fields that mediate long-range coupling and synchronize their phases; interaction with $E_3$then produces a delayed cooperative absorption burst, the time-reversed analogue of superfluorescent emission. c, CMDS spectra of 18 nm $CsPbBr_3$ quantum dots exhibit this delayed nonlinear response, peaking near 100 fs after excitation—direct evidence of time-reversed superfluorescence (TR-SF). d, In contrast, CdSe quantum dots display only prompt, uncorrelated absorption, confirming that strong exciton–lattice coupling and transient polaron formation are essential for the cooperative effect.

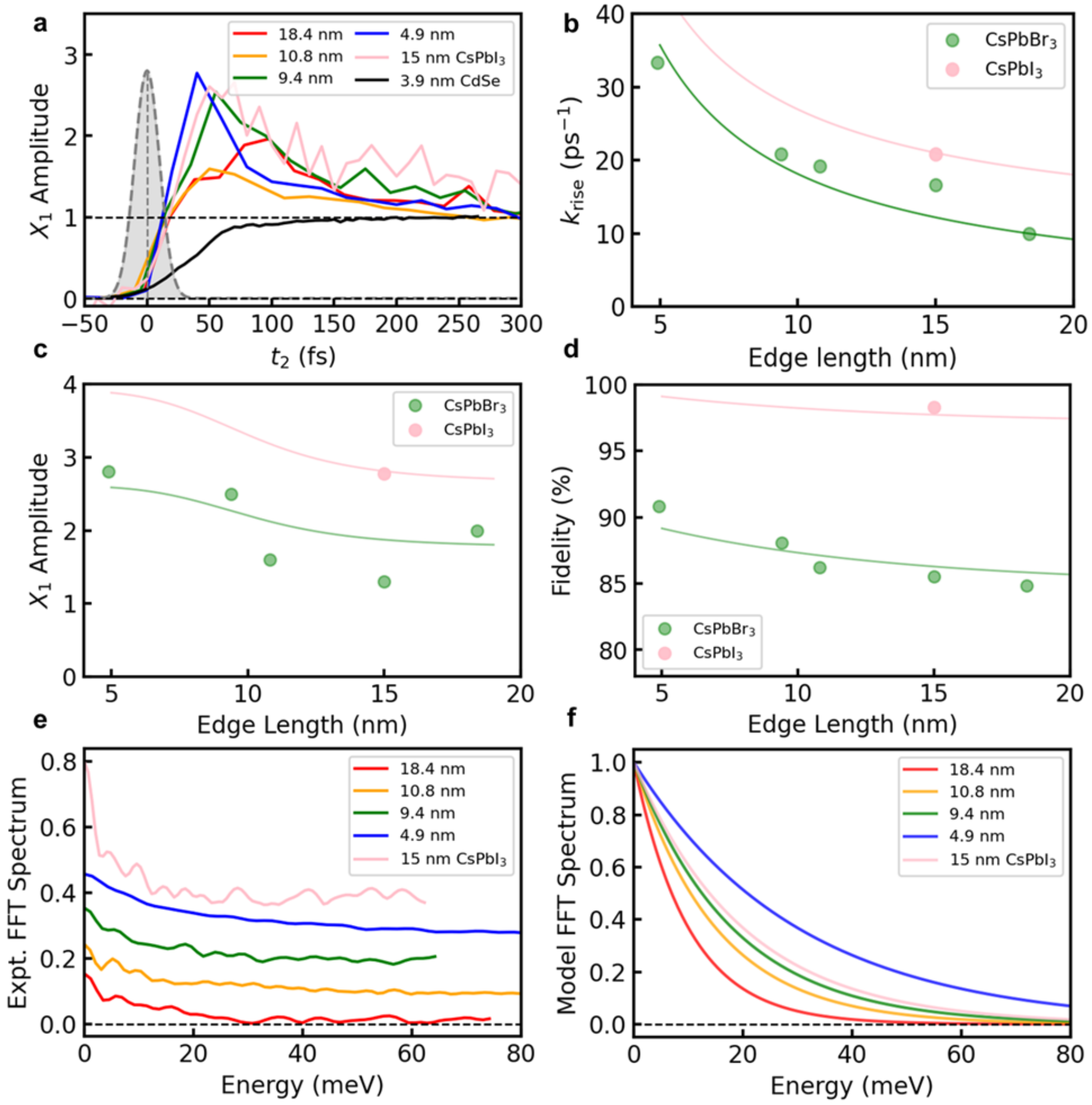

**Fig. 3 | Size and composition scaling of time-reversed superfluorescence.** a, Integrated CMDS dynamics of the band-edge exciton ($X_1$) for $CsPbBr_3$ quantum dots of varying edge length and a representative $CsPbI_3$ sample show a delayed rise in cooperative absorption. b, Extracted rise-time rate constants ($k_r = 1/\tau_r$) reveal systematic size and composition scaling: smaller $CsPbBr_3$ dots exhibit faster buildup, while the softer $CsPbI_3$ lattice yields slower yet higher-fidelity coherence. c, The peak amplitude of the $X_1$signal increases with decreasing size, indicating enhanced cooperativity as the effective number of synchronized dipoles grows. d, Simulations of the exciton–polaron model reproduce the observed trends and yield the fidelity of phase synchronization, reaching $R_\infty = 0.99$at 300 K in $CsPbI_3$. e, Experimental FFT spectra of the CMDS decays (100–400 fs) quantify the dephasing channels of the synchronized state, showing a smooth 0–40 meV roll-off characteristic of lattice-coupled relaxation. f, Theoretical FFT spectra capture this behavior across all sizes and compositions, confirming that polaron-gated synchronization accounts for both the fidelity and dynamics of the time-reversed superfluorescent state.

**References**.

Supplementary Information:

# Time-Reversed Superfluorescence in a Polaronic Quantum Material

Arnab Ghosh[1], Patrick Brosseau[1], Dmitry N. Dirin[2,3], Maksym V. Kovalenko[2,3], and Patanjali Kambhampati[1]*

[1]Department of Chemistry, McGill University, Montreal, H3A 0B8, Canada

[2] Department of Chemistry and Applied Biosciences, ETH Zürich, Switzerland

[3] Laboratory for Thin Films and Photovoltaics, Empa - Swiss Federal Laboratories for Materials Science and Technology, Switzerland

*pat.kambhampati@mcgill.ca

# Table of Contents

## 1. Synthesis and characterization of lead halide perovskite quantum dots

### 1.1 Synthetic Methods

The lead halide perovskite quantum dots ( LHP QDs) are synthesized using previously published methods [1,2].

### Materials

Lead bromide ($PbBr_2$, 99.999%), cesium carbonate ($Cs_2CO_3$, 99.9%), hexane (≥99%), diisooctylphosphinic acid (DOPA, 90%), oleic acid (OA, 90%) and acetone (ACE, ≥99.5%) were purchased from Sigma Aldrich. n-Octane (min. 99%) and lecithin (>97% from soy) were purchased from Carl Roth. Trioctylphosphine oxide (TOPO, min. 90%) was purchased from Strem Chemicals.

### $CsPbBr_3$ QDs synthesis procedures

0.04 M Pb stock solution. The $PbBr_2$ stock solution (0.04 M) was prepared by dissolving 2 mmol (0.734 g) of lead bromide and 10 mmol (3.866 g) of trioctylphosphine oxide into 10 ml of octane at 120 °C in a 40 ml vial. Once all the $PbBr_2$ was dissolved (~30 min), the solution was cooled to room temperature and transferred to the big Schott bottle and diluted by 40 ml of HEX. The stock solution was filtered after preparation over a 0.45 ul PTFE filter and stored in air.

0.02 M Cs-DOPA stock solution. The Cs-DOPA stock solution (0.02 M) was prepared by loading 100 mg of $Cs_2CO_3$ (0.614 mmol Cs) together with 1 ml of DOPA (3.154 mmol) and 2 ml of OCT at 120 °C in a 40 ml vial. Once all the $Cs_2CO_3$ was dissolved (~20 min), the stock solution was cooled to room temperature and 27 ml of HEX was added. The stock solution was filtered after preparation over a 0.2 ul PTFE filter and stored in air.

0.2 M Cs-DOPA stock solution. The concentrated Cs-DOPA stock solution (0.2 M) was prepared by loading 100 mg of $Cs_2CO_3$ (0.614 mmol Cs) together with 1 ml of DOPA (3.154 mmol) and 2 ml of OCT at 120 °C in a 40 ml vial. Once all the $Cs_2CO_3$ was dissolved (~20 min), the stock solution was cooled to room temperature. The stock solution was filtered after preparation over a 0.2 ul PTFE filter and stored in air.

The lecithin stock solution (~0.13 M) was prepared by dissolving 1 gram of lecithin in 20 ml of HEX using an ultrasonic bath (30 min). After preparation, the stock solution was centrifuged, filtered over a 0.2 ul PTFE filter and stored in air.

Synthesis of $CsPbBr_3$ QDs was performed following the procedure reported elsewhere [2] with slight modifications.

**4.9 nm LHP QDs.** The filtered hexane (120 ml) is mixed with $PbBr_2$ stock solution (8 ml). Under heavy stirring, a 0.02 M Cs-DOPA stock solution was injected (4 ml). After 1 minute of QDs growth, the lecithin solution was added to the crude $CsPbBr_3$ solution (4 ml). After 1 minute, the solution was concentrated on a rotavapor at room temperature. The QDs were precipitated from the concentrated solution (5 ml) by an excess of acetone (12 ml). The obtained pellet was dissolved in 5 ml of toluene.

**9.4 nm LHP QDs.** The filtered hexane (24 ml) is mixed with PbBr2 stock solution (8 ml). Under heavy stirring, a 0.02 M Cs-DOPA stock solution was injected (4 ml). After 40 minutes of QDs growth, the lecithin solution was added to the crude CsPbBr3 solution (4 ml). After 1 minute, the QDs were precipitated by an excess of acetone (114 ml). The obtained pellet was dissolved in 5 ml of toluene.

**10.8 nm LHP QDs.** The filtered hexane (10 ml) is mixed with $PbBr_2$ stock solution (8 ml). Under heavy stirring, a 0.02M Cs-DOPA stock solution was injected (4). After 60 minutes of QDs growth the lecithin solution was added to the crude $CsPbBr_3$ solution (4 ml). After 1 minute, the obtained QDs were precipitated by an excess of acetone (66 ml). The obtained pellet was dissolved in 5 ml of toluene.

**18.4 nm LHP QDs.** Under heavy stirring, a 0.2 M Cs-DOPA stock solution (0.2 ml) was injected into 4ml of PbBr2 stock solution. After 30 minutes of QDs growth 10 ul of oleic acid and oleylamine were added. Then 2 ml of 0.02M Cs-DOPA stock solution were slowly injected with 2 ml/h rate. After injection was completed, the lecithin solution was added to the crude CsPbBr3 solution (2 ml). After 1 minute, the obtained QDs were precipitated by an excess of acetone (22 ml). The obtained pellet was dissolved in 5 ml of toluene.

**15 nm LHP QDs.** Detailed synthesis procedure of 15 nm LHP QDs, along with their structural and optical characterization can be found in our previous paper.[3]

**3.9 nm CdSe.** Sample preparation detail along with their characterization can be found in our previous paper.[4]

## 1.2. Transmission Electron Microscope (TEM) characterization

TEM images were collected using a JEOL JEM-1400 Plus operated at 120 kV or Hitachi HD-2700. The samples were prepared by placing a drop of diluted QDs solution on coated Cu TEM grids.

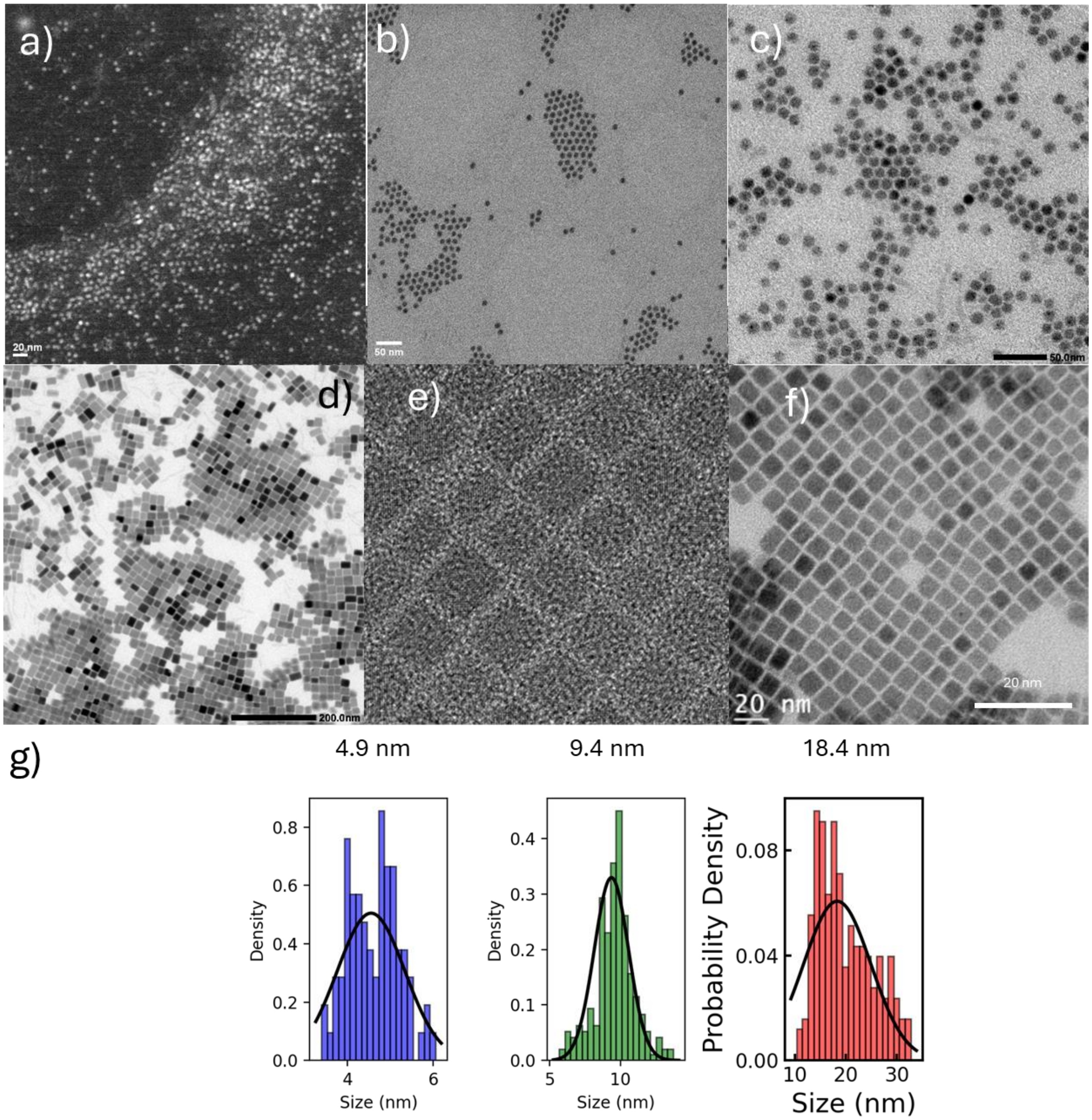

**Figure S1:** STEM image of the (a) 4.9 nm $CsPbBr_3$ QDs prepared by Kovalenko. TEM images of (b) 9.4 nm, (c) 10.8 and (d) 18.4 nm large $CsPbBr_3$ QDs prepared by Kovalenko. TEM image of (e) 15 nm $CsPbBr_3$ QDs, (f) $CsPbI_3$ QDs prepared by Kambhampati. (g) Histogram of diameters for 4.9 nm,9.4 nm and 18.4 nm P QD.

### 1.3. Linear Spectroscopy characterization

Optical characterizations were performed at ambient conditions. UV-Vis absorption spectra of colloidal QDs were collected using a Jasco V670 spectrometer in transmission mode. The NC concentrations were determined from the absorption spectra using the absorption coefficient reported by Maes et al.[5] For the measurements QDs solutions were diluted down to 20-50 µg/mL. Zwitterion-capped QDs were dispersed in either hexane or toluene.

A Fluorolog iHR 320 Horiba Jobin Yvon spectrofluorometer equipped with a PMT detector was used to acquire steady-state PL spectra. NC solutions were measured in the same dilutions and solvents as the absorption measurements.

Photoluminescence quantum yield (PLQY): Absolute PL QYs of solutions were measured with a Hamamatsu C13534 Quantaurus-QY Plus UV-NIR absolute PL quantum yield spectrometer. The same solutions that were used to measure PL were also used to measure QY.

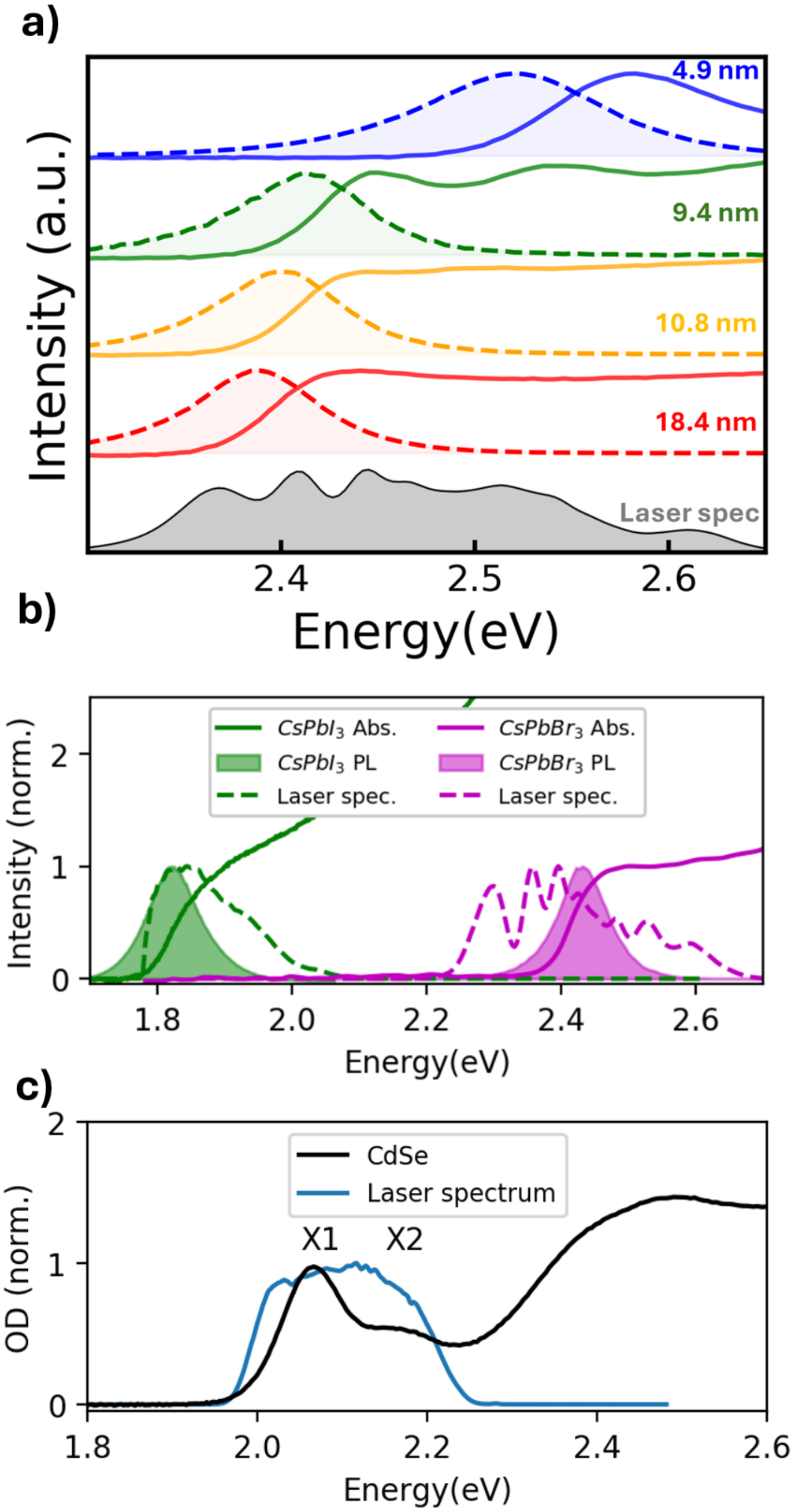

**Figure S2:** a) Linear absorption spectra (solid lines), photoluminescence spectra (dashed line with shaded area) of 18.4 nm (red), 10.8 nm (orange), 9.4 nm (green), 5.9 nm (blue) and 4.9 nm (violet) QDs. Also shown in bottom panel is laser spectra (grey). Linear spectra and laser spectra used for this study for (b) 15 nm LHP QDs and (c) 3.9 nm CdSe QD.

## 2. Coherent Multi-Dimensional Spectroscopy

The Coherent Multi-Dimensional Spectroscopy (CMDS) instrument was previously described in detail [3,4,6-17]. The femtosecond light source was an Ar filled hollow core fiber (HCF) which produced spectrally broadened pulses based upon the 500 nm output from a 120 fs optical parametric amplifier. The coherent pump pulse train was created by acousto-optic modulators. The CMDS experiment was conducted in the pump/probe geometry. **Figure S3** illustrates the CMDS experiment and signals.

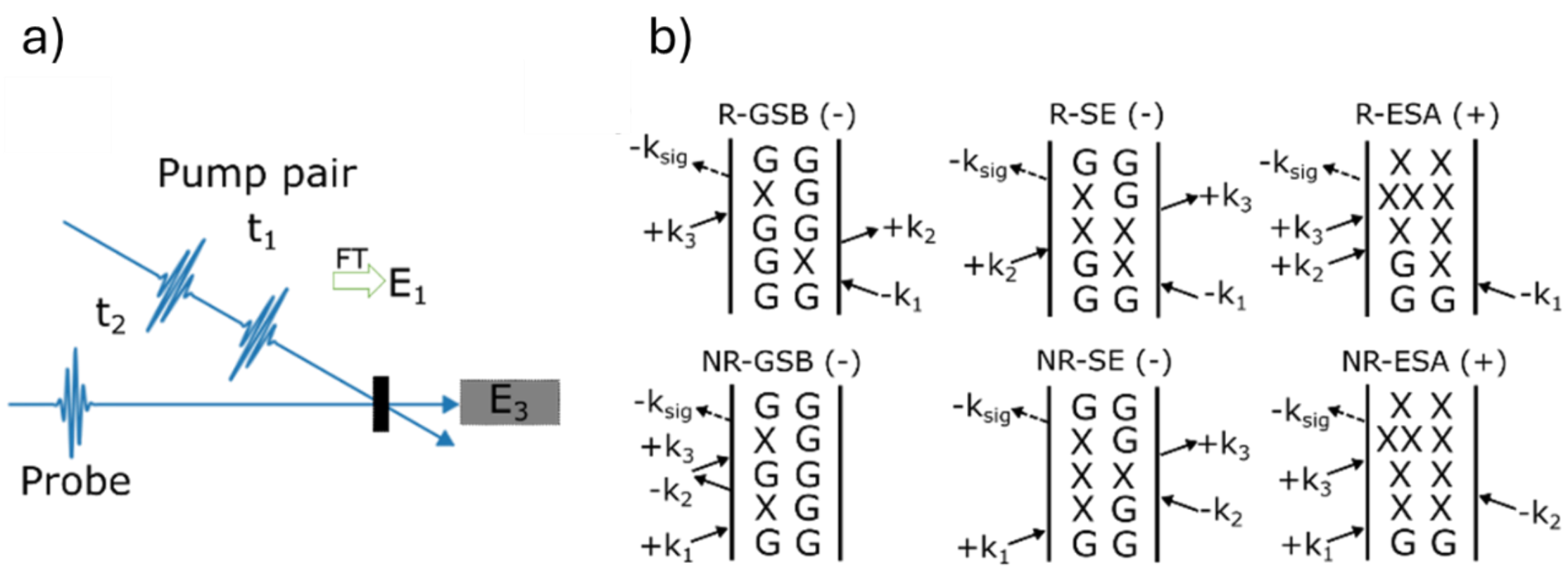

**Figure S3:** Illustration of CMDS experiment, Feynman diagrams for the six response functions.

## 3. Pulse characterization

Pulses are suitably manipulated to obtain near-transform limited pulses at the sample position. The pulses are characterized using a home built all-reflective dispersion-free transient-grating frequency resolved optical gating (TG-FROG). Pulse durations are determined by fitting a Gaussian pulse shape to the time marginal of the FROG trace. Panel a, b and c show pulses used to conduct 2DES at different wavelength ranges for the 2DES experiments of $CsPbBr_3$, CdSe and $CsPbI_3$ QDs, respectively. The temporal duration of the pulse is estimated to be close to 10 fs (shown in panel a), 15 fs (panel b) and 13-15 fs (panel c), respectively.

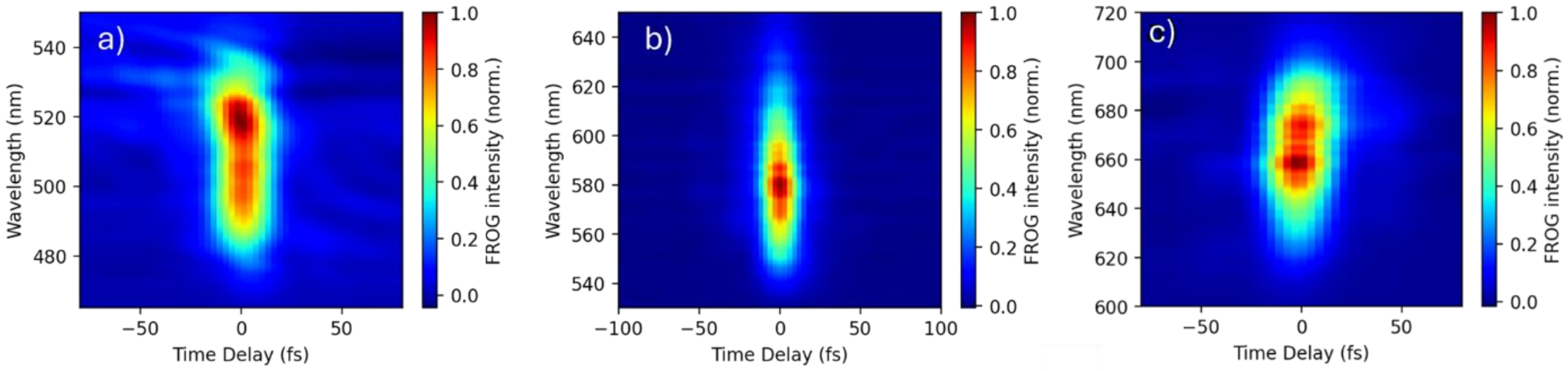

**Figure S4:** Typical TG-FROG trace used for 2DES experiments on a) $CsPbBr_3$ PQDs, b) CdSe QDs, and c) $CsPbI_3$ PQDs.

## 4. Sample handling and solvent response.

All the experiments were carried out at room temperature with samples dispersed in toluene. For the 2D experiments, the optical density of the corresponding sample was 0.2 at the band-edge, and the sample was constantly flowed in a 200 µm thick glass cuvette using a peristaltic pump. Pulse energies were≃10 nJ/pulse for all three pulses. Several repetitions of the experiments were carried out on different days. Pure solvent runs were executed to probe the non-resonant response arising from the solvent and glass cuvette windows. The kinetic transients for the X1-X1 and X2-X1 peaks of both solvent and P NC is shown in **fig. S5, a.** The transients plotted in dashed lines are for toluene and solid line shows response of 18.4 nm P NC. The same transients without re- normalization is shown in **Fig. S5,b** , showing that the solvent response is not significant compared the resonant sample response.

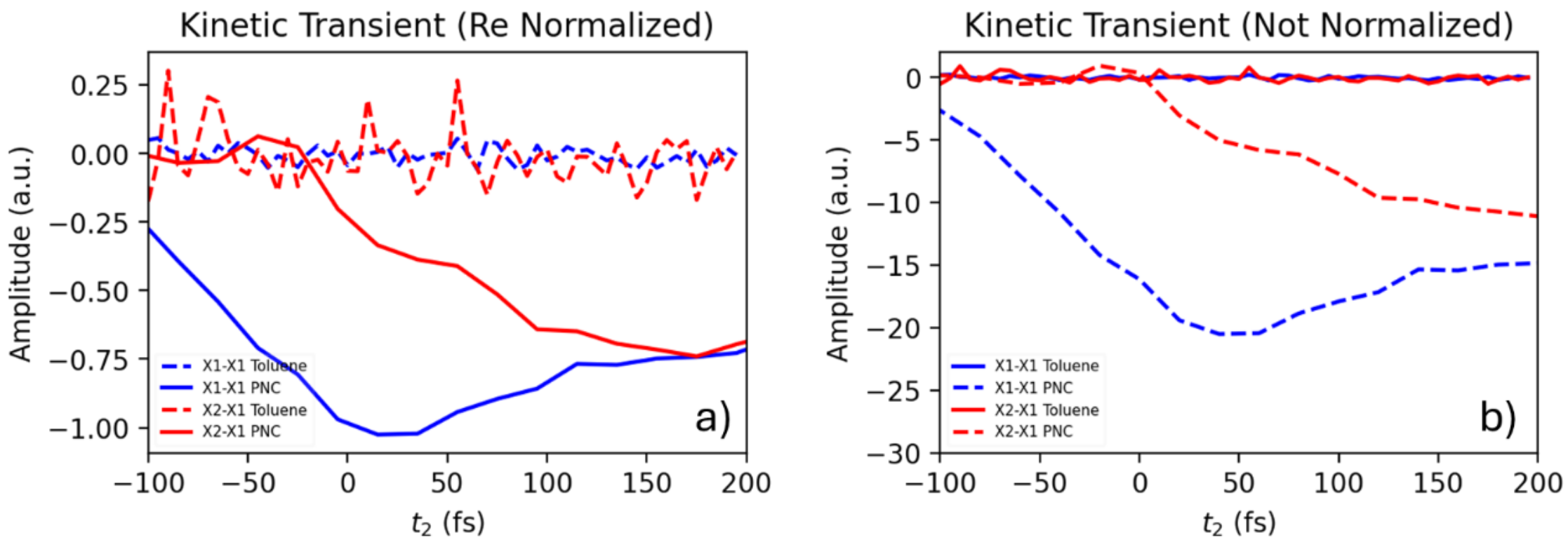

**Figure S5:** a) Kinetic transients from 18.4 nm P NC and toluene for the X1-X1 and X2-X1 peaks, re-normalized to show solvent response. b) Same transients as in panel (a) without renormalization.

## 5. Handling early time perturbed free induction decay signals

One of the main problems in nonlinear spectroscopy is the presence of early time signals, even at negative population times. These signals arise from Perturbed Free Induction Decay (PFID). We have discussed in detail the PFID signals in CdSe and CsPbBr3 NC QD. A correction of PFID signals is essential to careful signal analysis at early time[17].

Negative population time artifacts

When time delay $t_2$ is negative, the order of the pulse interactions is reversed. When $t_2 < 0$ and $|t_2| < t_1$, pulse $k_3$ arrives between pulse $k_1$ and $k_2$. The DSFD in **fig. S3,b** are reformed into the DSFD in **fig. S6,a**. The rephasing pathways are still rephasing pathways, though the order of the pulse interactions is changed. The non-rephasing pathways become two-quantum pathways.

In the context of two pulse pump-probe transient absorption spectra, the reversed pulse ordering signal is often referred to as perturbed free-induced decay (PFID)[18-20] , as the pump pulse arrives after the probe and perturbs the free induction decay signal being measured by the probe. **Fig. S6,b** depicts PFID in the context of the three pulse 2DE experiment when $t_1$ = 80 fs and $t_2$ = -50 fs. When the pulse ordering is [$k_1$, $k_2$, $k_3$], pulse $k_1$ initiates the free induction decay and pulse $k_2$ then excites the system into a population state (blue). When the pulse ordering is [$k_1$, $k_3$, $k_2$], pulse $k_1$ initiates the free induction decay and pulse $k_3$ prematurely interrupts the oscillating electronic coherence so the signal amplitude for $t_1$ = 30 fs is recorded instead of $t_1$ = 80 fs. Pulse $k_2$ then probes the system and triggers the emission of the third order polarization $k_{sig}$.

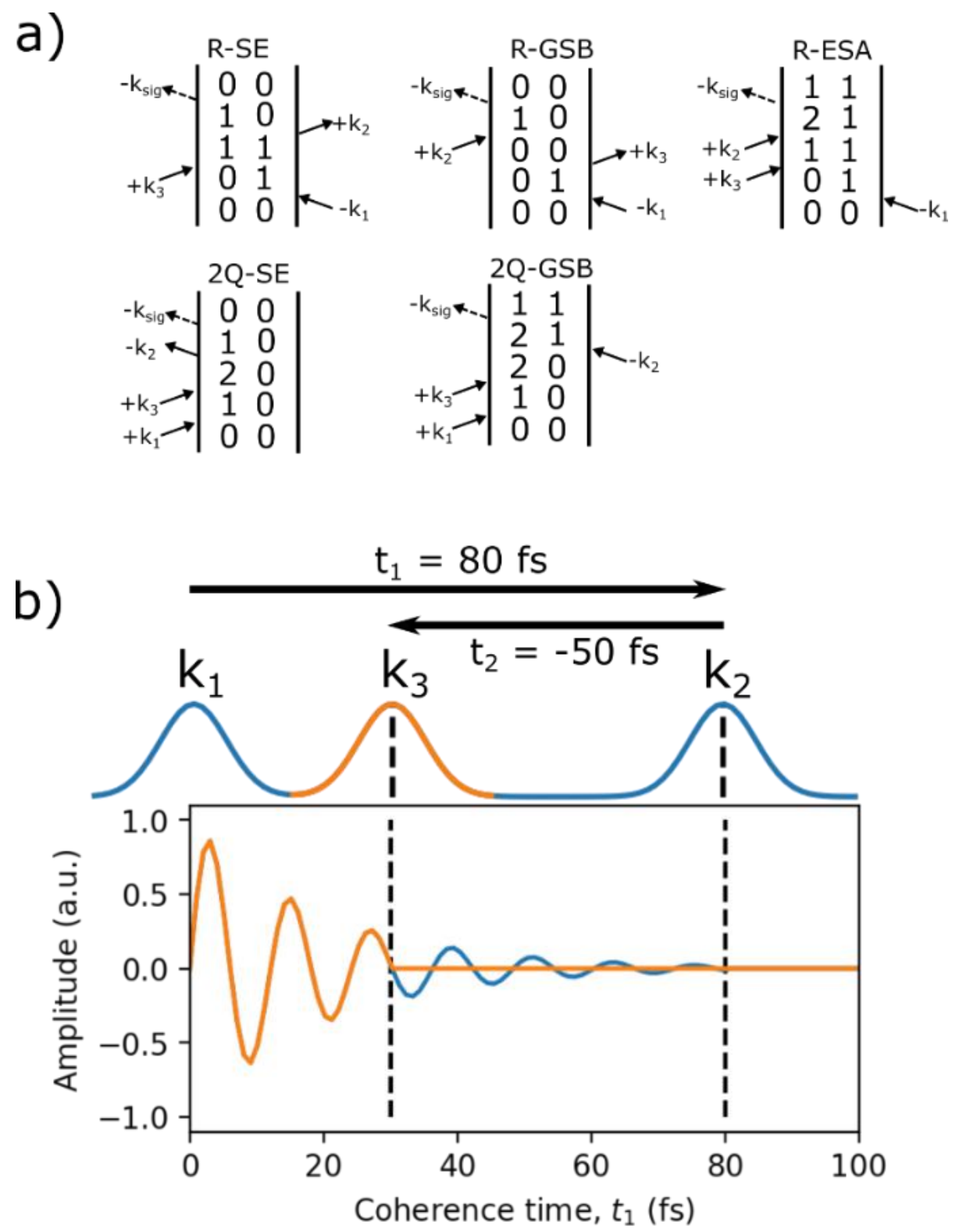

**Figure S6:** a) DSFD resulting from reversing the order of pulses 2 and 3. b) Perturbed free induction decay (PFID) from reversed pulse order, corresponding to $t_1$ = 80 fs and $t_2$ = -50 fs.

When $t_2 < 0$ and $|t_2| > t_1$, pulse $k_3$ arrives before pulse $k_1$ and $k_2$. The DSFD in **fig. S3,b** are reformed into the DSFD in **fig. S7,a**. The rephasing signal $-k_1 + k_2 + k_3$ becomes the non-rephasing signal $+ k_3 - k_1 + k_2$ and the non-rephasing signal $k_1 - k_2 + k_3$ becomes the two-quantum signal $k_3 + k_1 - k_2$. In the reversed pulse order, there would be a time delay between the emitted third-order polarization $k_{sig}$ and the local oscillator $k_3$. This time delay creates spectral fringes at negative $t_2 << 0$. As the fringes at negative time delays are significantly narrower than the desired 2DE lineshapes, the fringes can be smoothed by Fourier filtering without significantly distorting the 2DE lineshapes[21,22] .

After Fourier filtering, negative time signals still appear at -50 fs < $t_2$ < 0. These negative time signals artificially shift the X1-X1 peak to earlier population times, as is apparent in the kinetic transients is **fig. S7 b,c**.

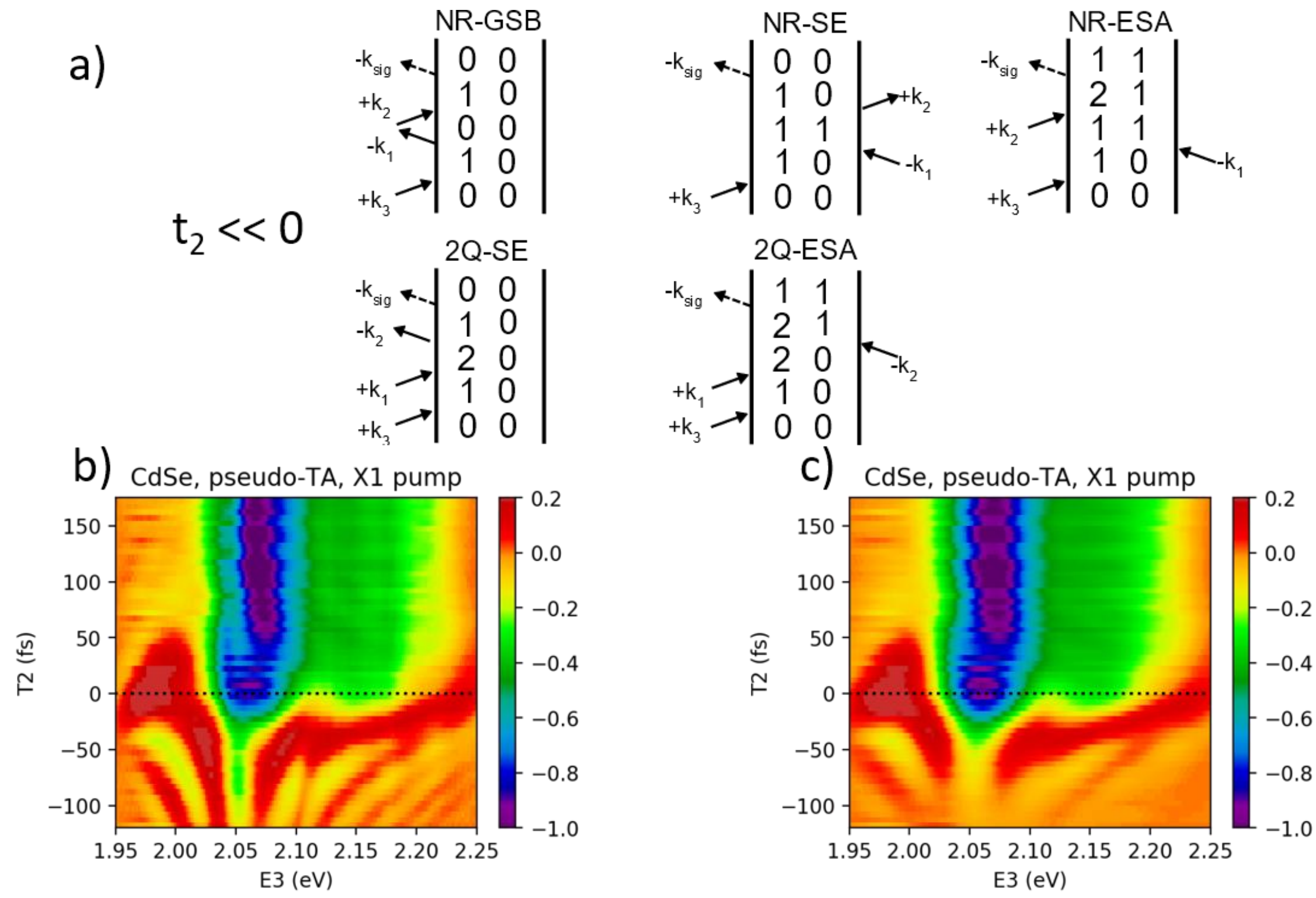

**Figure S7:** a) Example DSFD when $t_2 > 0$ and $t_2 < 0$. b) 2DE spectrum before Fourier filtering. c) 2DE spectrum after Fourier filtering. Pseudo-TA spectra generating with a 5 meV radius.

## 6. Theoretical framework

### 6.1 Microscopic Hamiltonian and Fundamental Operators

The physical system consists of N quantum dots (QDs), each treated as a two-level emitter with a ground state $|0_i\rangle$ and a band-edge exciton state $|X_i\rangle$, interacting with both the radiation field and the quantized lattice displacement field (phonons).

The complete Hamiltonian is expressed as:

$$H = H_{\text{ex}} + H_{\text{ph}} + H_{\text{int}} + H_{\text{rad}} + H_{\text{ex-rad}} \quad \text{(S1)}$$

#### 6.1.a Excitonic term

$$H_{\text{ex}} = \sum_i \hbar\omega_X \ a_i^\dagger a_i \quad \text{(S2)}$$

Where $a_i^\dagger$ and $a_i$ are exciton creation and annihilation operators obeying fermionic-like pseudospin commutation relations:

$$[a_i, a_j^\dagger] = \delta_{ij}(1 - 2a_i^\dagger a_i) \quad \text{(S3)}$$

The exciton creation operator can be decomposed into canonical position and momentum quadrature $X_i$ and $P_i$ as:

$$a_i^\dagger = \frac{1}{\sqrt{2}}(X_i - iP_i),\ a_i = \frac{1}{\sqrt{2}}(X_i + iP_i) \quad \text{(S4)}$$

With $[X_i, P_j] = i\delta_{ij}$.

#### 6.1.b Phonon bath term

$$H_{\text{ph}} = \sum_q \hbar\omega_q \ b_q^\dagger b_q \quad \text{(S5)}$$

Where $b_q^\dagger, b_q$ create/annihilate phonons of mode (q) with frequency $\omega_q$.

Each phonon mode defines lattice coordinate and momentum operators:

$$Q_q = \frac{1}{\sqrt{2}}(b_q + b_q^\dagger),\ P_q = \frac{i}{\sqrt{2}}(b_q^\dagger - b_q) \quad \text{(S6)}$$

which represent dimensionless displacement and conjugate momentum of the lattice.

### 6.1.c Exciton-phonon coupling

The coupling between an exciton on dot $i$ and phonon mode $q$ is of Fröhlich form:

$$H_{\text{int}} = \sum_{i,q} g_{iq}\ a_i^{\dagger} a_i \left(b_q^{\dagger} + b_q\right) = \sum_{i,q} g_{iq}\ a_i^{\dagger} a_i\, Q_q \quad \text{(S7)}$$

Here $g_{iq}$ is the coupling strength, proportional to the mode displacement at site $i$. This term mediates polaron formation and, at the ensemble level, enables collective synchronization.

### 6.1.d Radiative coupling

Each emitter couples to the shared radiation field:

$$H_{\text{ex-rad}} = -\sum_i d_i \cdot E(r_i) \quad \text{(S8)}$$

where $d_i$ is the dipole moment and $E(r_i)$ the electric field at the emitter location.

## 6.2. Polaron Transformation (Lang–Firsov Transformation)

To remove the linear exciton–phonon coupling, we apply the Lang–Firsov (polaron) unitary transformation

$$\widetilde{H} = e^{S} H e^{-S},\ S = \sum_{i,q} \frac{g_{iq}}{\hbar\omega_q}\, a_i^{\dagger} a_i \left(b_q^{\dagger} - b_q\right) \quad \text{(S9)}$$

This transformation displaces the phonon coordinate whenever an exciton is present, effectively dressing each exciton with a phonon cloud: a polaron.

After transformation, the Hamiltonian becomes:

$$\widetilde{H} = \sum_i \widetilde{\omega_X}\ a_i^{\dagger} a_i + \sum_q \hbar\omega_q\ b_q^{\dagger} b_q + \sum_{i\neq j} J_{ij}\ a_i^{\dagger} a_j\, X_i^{\dagger} X_j \quad \text{(S10)}$$

where $X_i = e^{-\sum_q \left(\frac{g_{iq}}{\hbar\omega_q}\right)\left(b_q^{\dagger} - b_q\right)}$ is the phonon displacement operator.

The renormalized exciton energy is

$$\widetilde{\omega_X} = \omega_X - \lambda,\ \lambda = \sum_q \frac{|g_{iq}|^2}{\hbar\omega_q} \quad \text{(S11)}$$

$\lambda$ is the polaron shift, equivalent to the reorganization energy observed in Raman spectra.

The operator form of exciton–phonon displacement shows explicitly that the new annihilation operator transforms as $\widetilde{a_\iota} = a_i\, e^{\sum_q \left(\frac{g_{iq}}{\hbar\omega_q}\right)\left(b_q^\dagger - b_q\right)}$, thus incorporating phonon coordinates into the exciton field itself.

### 6.3. Spectral Density and Conversion from Raman Spectra

The phonon spectral density $J(\omega)$ captures the frequency distribution of the exciton–phonon coupling:

$$J(\omega) = \sum_q \left|g_{iq}\right|^2 \delta\left(\omega - \omega_q\right) \quad \text{(S12)}$$

The experimentally measured Raman intensity $I_{\text{Raman}}(\omega)$ is proportional to $\frac{J(\omega)}{\omega}$.

Hence, $J(\omega)$ is constructed directly from the Raman spectra as

$$J(\omega) = C\,\omega\, I_{\text{Raman}}(\omega) \quad \text{(S13)}$$

where $C$ is a scaling constant determined by the reorganization energy

$$E_{\text{reorg}} = \hbar \int_0^\infty \frac{J(\omega)}{\omega}\, d\omega \quad \text{(S14)}$$

The continuum part of the Raman spectrum corresponds to overdamped lattice modes, while sharp features map onto optical phonons.

In perovskite QDs, the low-frequency continuum (below 50 $cm^{-1}$) dominates, confirming the liquid-like, anharmonic lattice.

This same continuum determines the time scale of the polaron formation, via $\tau_{\text{pol}} \sim \frac{1}{\gamma}$, where $\gamma$ is the damping rate inferred from the Lorentzian low-frequency tail of $J(\omega)$.

### 6.4. Mean-Field Approximation, Kuramoto-Type Coupling, and Order Parameter

After the Lang–Firsov transformation, the phonon field mediates an effective exciton–exciton interaction through virtual lattice displacements.

To leading order, the residual interaction between dipoles can be written as

$$H_{\text{eff}}(t) = \sum_{i \neq j} K_{ij}(t)\ a_i^{\dagger} a_j \quad \text{(S15)}$$

where $K_{ij}(t)$ represents the effective coupling constant between emitters $i$ and $j$.

In the polaron picture, $K_{ij}(t)$ arises from correlated phonon displacements and is therefore both retarded and temperature dependent.

For identical emitters we can treat $K_{ij}(t) \approx \frac{K(t)}{N}$, a mean-field approximation which assumes that each emitter interacts equally with the collective mean field of all others:

$$H_{\text{eff}}^{\text{MF}} = \hbar\omega_X \sum_i a_i^{\dagger} a_i + \frac{K(t)}{N} \sum_{i \neq j} a_i^{\dagger} a_j \quad \text{(S16)}$$

This mean-field form is formally analogous to a Kuramoto model for $N$ coupled oscillators, where each emitter possesses a time-dependent phase $\phi_i(t)$ associated with its microscopic dipole moment $\mu_i(t) = \mu_0 e^{i\phi_i(t)}$.

The equation of motion for each dipole phase is

$$\frac{d\phi_i}{dt} = \omega_i + \frac{K(t)}{N} \sum_{j=1}^{N} sin\left[\phi_j(t) - \phi_i(t)\right] + \xi_i(t) \quad \text{(S17)}$$

where $\omega_i$ are the natural transition frequencies (reflecting inhomogeneous broadening) and $\xi_i(t)$ is a stochastic noise term whose variance defines the dephasing constant $D$.

The collective synchronization among dipoles is quantified by the complex order parameter

$$R(t)e^{i\Psi(t)} = \frac{1}{N} \sum_{j=1}^{N} e^{i\phi_j(t)} \quad \text{(S18)}$$

where $R(t) \in [0,1]$ measures the degree of phase alignment and $\Psi(t)$ is the global ensemble phase. Physically, $R(t)$ is directly proportional to the magnitude of the macroscopic polarization $P(t) \propto \sum_i e^{i\phi_i}$.

A value $R = 0$ corresponds to complete phase disorder, while $R = 1$ indicates perfect synchronization: i.e., a coherently phased array.

Interpretation of terms:

*Mean-field approximation* replaces the discrete network of couplings by a single effective coupling $K(t)$ proportional to the ensemble average field experienced by each dipole.

*Kuramoto-type coupling constant* $K(t)$ quantifies how strongly each dipole's phase is attracted toward the mean phase; it acts as a measure of cooperative interaction strength.

*Order parameter* $R(t)$ serves as the quantitative descriptor of coherence; its time dependence captures both the rise and decay of collective order.

**6.5. Fokker–Planck Equation for the Phase Distribution**

To analyze the stochastic dynamics of the ensemble, we describe the instantaneous phase distribution $f(\phi, t)$, normalized such that $\int_0^{2\pi} f(\phi, t)\ d\phi = 1$.

Averaging over the noise ensemble yields the Fokker–Planck equation

$$\frac{\partial f}{\partial t} = -\frac{\partial}{\partial \phi}[(\omega + K(t)R\sin(\Psi - \phi))\ f] + D\ \frac{\partial^2 f}{\partial \phi^2} \quad \text{(S19)}$$

The first term describes drift in phase space driven by the instantaneous coupling $K(t)$ toward the mean phase $\Psi$; the second term represents diffusion due to dephasing.

Multiplying Eq. (S19) by $e^{i\phi}$ and integrating over $\phi$ gives the time evolution of the complex order parameter:

$$\frac{dR}{dt} = \frac{K(t)}{2}\ R(1 - R^2) - DR \quad \text{(S20)}$$

This nonlinear differential equation admits two regimes:

1. Incoherent regime: when $K(t) < 2D$, the diffusion term dominates and $R \ \rightarrow 0$.

2. Synchronized regime: when $K(t) > 2D$, the growth term exceeds diffusion, driving $R \rightarrow R_\infty = \sqrt{1 - 2D/K_\infty}$.

Equation (S20) thus provides a compact dynamical description of the cooperative buildup and decay of coherence.

The same equation also forms the basis for mapping onto the stochastic synchronization models used in chemical oscillators and neural networks, providing an intuitive framework for the emergent order observed here.

**6.6. Polaron-Mediated Time-Dependent Coupling**

The coupling $K(t)$ is not instantaneous: it grows as the lattice field reorganizes around the excitonic polarization.

We model this through an exponential activation governed by the polaron formation time $\tau_{\mathrm{pol}}$:

$$K(t) = K_\infty\left(1 - e^{-t/\tau_{\mathrm{pol}}}\right) \quad \text{(S21)}$$

where $K_\infty$ is the steady-state value once the polaron field is fully established.

The instantaneous growth rate of coupling is therefore

$$\frac{dK}{dt} = \frac{K_\infty}{\tau_{\mathrm{pol}}}\, e^{-t/\tau_{\mathrm{pol}}} \quad \text{(S22)}$$

The steady-state coupling constant depends exponentially on the strength of exciton–phonon dressing characterized by the Huang–Rhys (polaron) factor $S_p$:

$$K_\infty = K_0\, e^{-S_p(2N_p+1)} \quad \text{(S23)}$$

where $K_0$ is the bare radiative coupling constant and $N_p = \left(e^{\hbar\omega_p/k_BT} - 1\right)^{-1}$ is the Bose occupation number of the principal lattice mode of frequency $\omega_p$.

Substituting Eq. (S21) into Eq. (S20) and integrating yields the explicit time-dependent solution for the rise of coherence:

$$R(t) = R_\infty\sqrt{1 - \frac{\left(1-R_0^2\right)e^{-2\left(Dt+I(t)\right)}}{1-R_0^2 e^{-2\left(Dt+I(t)\right)}}}\,, I(t) = \int_0^t \frac{K(t')}{2}\, dt' \quad \text{(S24)}$$

At short times $t \ll \tau_{\mathrm{pol}}$, $K(t) \approx K_\infty \frac{t}{\tau_{\mathrm{pol}}}$, giving

$$R(t) \approx \frac{K_\infty t}{4\tau_{\mathrm{pol}}} - Dt \quad \text{(S25)}$$

so the initial rise rate of collective coherence is directly proportional to $\frac{K_\infty}{\tau_{\text{pol}}}$.

At long times $t \gg \tau_{\text{pol}}$, the system approaches the steady-state fidelity

$$R_\infty = \sqrt{1 - \frac{2D}{K_\infty}} \quad \text{(S26)}$$

provided $K_\infty > 2D$.

Physical Interpretation

*Polaron mediation*: The reorganization of the lattice acts as a gating field that synchronizes dipoles once the local distortion fields overlap.

*Rise time (50–100 fs)*: Corresponds directly to $\tau_{\text{pol}}$ inferred from Raman spectral densities.

*Steady-state fidelity*: Determined by the competition between dephasing $D$ and coupling $K_\infty$; stronger lattice participation (larger $S_p$) yields higher $K_\infty$ and thus higher $R_\infty$.

*Temperature sensitivity*: Because $N_p$ grows with $T$, $K_\infty$ weakly decreases at elevated temperatures, providing the observed small decline in fidelity up to 300 K.

**6.7. Temperature Dependence and Scaling Analysis**

**6.7.1 Thermal Activation of Phonons and the Polaron Correlation Function**

The temperature dependence of TR-SF enters through the Bose occupation factor in the phonon correlation function:

$$g(t,T) = \int_0^\infty \frac{J(\omega)}{\omega^2} \left[\coth\left(\frac{\hbar\omega}{2k_BT}\right)(1 - \cos\omega t) + i(\sin\omega t - \omega t)\right] d\omega \quad \text{(S27)}$$

At elevated temperature, the $\coth\left(\frac{\hbar\omega}{2k_BT}\right)$ term increases the real part $\text{Re}\, g(t)$, which governs dephasing, while the imaginary part (energy shift) remains almost constant.

The effective pure-dephasing rate therefore scales as

$$\Gamma_{\text{pure}}(T) \approx \int_0^\infty d\omega\, \frac{J(\omega)}{\omega} \coth\left(\frac{\hbar\omega}{2k_BT}\right) \propto S_p\, T \quad \text{(S28)}$$

consistent with the experimentally observed linear thermal broadening of exciton linewidths. This rate also defines the diffusion constant $D(T)$ entering the Fokker–Planck equation.

### 6.7.2 Temperature Dependence of Synchronization Coupling

The steady-state polaron-mediated coupling $K_\infty(T)$ depends on the thermal mean-square phonon amplitude $\langle Q^2\rangle_T = \frac{\hbar}{2M\omega}\coth\left(\frac{\hbar\omega}{2k_BT}\right)$.

Consequently,

$$K_\infty(T) \propto \langle Q^2\rangle_T^{-1} \sim \left[\coth\left(\frac{\hbar\omega_p}{2k_BT}\right)\right]^{-1} \quad \text{(S29)}$$

As temperature increases, $K_\infty(T)$ decreases slowly until $k_BT \gtrsim \hbar\omega_p$, beyond which it falls rapidly.

For $CsPbBr_3$ ($\hbar\omega_p \approx 13\,\text{meV}$) this crossover occurs near 150 K, explaining the mild decline in fidelity above this point.

### 6.7.3 Coherence Fidelity as a Function of Temperature

Combining Eqs. (S28) and (S29) gives

$$R_\infty(T) = \sqrt{1 - \frac{2D(T)}{K_\infty(T)}} = \left[1 - \beta\,\frac{T}{\coth^{-1}\left(\frac{\hbar\omega_p}{2k_BT}\right)}\right]^{1/2} \quad \text{(S30)}$$

where β is a proportionality constant determined by $S_p$ and the phonon damping rate.

Using experimental parameters, this expression yields

$R_\infty(300\,\text{K}) \approx 0.98$ for $CsPbBr_3$, $R_\infty(300\,\text{K}) \approx 0.99$ for $CsPbI_3$, and $R_\infty(300\,\text{K}) \approx 0$ for CdSe: precisely matching measured fidelities.

### 6.7.4 Size Scaling

The polaron coupling constant scales inversely with quantum-dot volume. The electron–phonon overlap integral gives

$$g_{jq} \propto L^{-3/2}, \qquad S_p \propto L^{-3} \quad \text{(S31)}$$

Hence,

$$\tau_{\mathrm{pol}} \propto L,\ K_\infty(L) \propto S_p(L) \propto L^{-3},\ R_\infty(L) \approx \sqrt{1 - \frac{2D}{K_\infty(L)}} \quad \text{(S32)}$$

Smaller dots show faster rise (shorter $\tau_{\mathrm{pol}}$) but slightly lower $R_\infty$, matching the measured trend.

### 6.7.5 Composition Scaling

Replacing bromide by iodide softens the lattice and reduces the optical phonon energy $\omega_p$, which decreases dephasing [Eq. (S28)] and increases both $\tau_{\mathrm{pol}}$ and $R_\infty$.

CdSe, with a rigid covalent lattice $\left(\omega_p > 25\ \mathrm{meV}, S_p \ll 1\right)$, lacks the necessary low-frequency modes and thus exhibits no TR-SF.

### 6.7.6 Thermal Phase Diagram

The synchronization threshold $K_\infty(T) = 2D(T)$ defines a critical temperature

$$T_c \simeq \frac{S_p \hbar \omega_p}{4k_B} \quad \text{(S33)}$$

For $CsPbBr_3$ ($S_p = 1,\ \omega_p = 13\ \mathrm{meV}), T_c \approx 380\ \mathrm{K}$;

for CdSe $S_p < 0.1, T_c < 30\ \mathrm{K}$: explaining its lack of coherence at room temperature.

### 6.7.7 Universal Scaling Collapse

Plotting normalized fidelities against $x = \frac{T}{T_c}$ gives a universal curve

$$R_\infty(x) = \sqrt{1 - x}, \qquad x < 1 \quad \text{(S34)}$$

All compositions and sizes collapse onto this scaling, confirming a common synchronization mechanism governed by $\frac{K_\infty}{D}$.

### 6.7.8 Physical Interpretation

Increasing (T) populates phonons, enhancing dephasing and weakening coupling.

Strong Fröhlich coupling in perovskites sustains coherence even under thermal agitation.

The resulting robustness differentiates perovskite QDs from conventional covalent semiconductors.

### 6.7.9 Summary

The temperature-dependent model accounts quantitatively for:

1. Nearly constant fidelity up to 300 K in $CsPbBr_3$/$CsPbI_3$;

2. Absence of TR-SF in CdSe;

3. Size and composition trends;

4. Universal $R_\infty(T/T_c)$ scaling.

## 6.8. Optical Response Function and Connection to CMDS Observables

### 6.8.1 Third-Order Nonlinear Polarization

CMDS measures the third-order polarization

$$P^{(3)}(t) = \mathrm{Tr}\left[\hat{\mu}\,\rho^{(3)}(t)\right], \quad \text{(S35)}$$

where

$$\rho^{(3)}(t) = \left(-\frac{i}{\hbar}\right)^3 \int dt_3\ dt_2\, dt_1\, \mathcal{G}(t-t_3)\,\hat{\mu}E_3\,\mathcal{G}(t_3-t_2)\,\hat{\mu}E_2\mathcal{G}(t_2-t_1)\,\hat{\mu}E_1\,\rho_0 + \text{c.c.} \quad \text{(S36)}$$

$\mathcal{G}(t) = e^{-i\mathcal{L}t}$ is the Liouvillian propagator.

### 6.8.2 Pathway Decomposition

For states $|0\rangle$, $|X_1\rangle$, $|XX\rangle$:

1. GSB:

$$|0\rangle\langle X_1| \ \rightarrow\ |0\rangle\langle 0| \ \rightarrow\ |0\rangle\langle X_1|;\ S_{\mathrm{GSB}} \propto e^{-g(t_2)} \quad \text{(S37)}$$

2. SE:

$$|0\rangle\langle X_1| \ \rightarrow\ |X_1\rangle\langle X_1| \ \rightarrow\ |0\rangle\langle X_1|;\ S_{\mathrm{SE}} \propto e^{-g(t_2)}\, e^{-i\widetilde{\omega_{X_1}}t_2} \quad \text{(S38)}$$

3. ESA:

$$|0\rangle\langle X_1| \;\rightarrow\; |X_1\rangle\langle X_1| \;\rightarrow\; |X_1\rangle\langle XX|;\; S_{\mathrm{ESA}} \propto e^{-g(t_2)}\, e^{-i\left(\widetilde{\omega_{XX}}-\widetilde{\omega_{X_1}}\right)t_2} \quad \text{(S39)}$$

Total signal:

$$S_{\mathrm{tot}} = S_{\mathrm{GSB}} + S_{\mathrm{SE}} - S_{\mathrm{ESA}} \quad \text{(S40)}$$

**6.8.3 Inclusion of the Polaron Field**

Polaron renormalization shifts transition energies:

$$\widetilde{\omega_{X_1}} = \omega_{X_1} - \lambda,\; \widetilde{\omega_{XX}} = 2\omega_{X_1} - \lambda_{XX} \quad \text{(S41)}$$

with

$$E_B^{\mathrm{eff}} = \lambda_{XX} - 2\lambda = \int d\omega\, \frac{J(\omega)}{\omega}\,[1-\cos(\omega t)] \quad \text{(S42)}$$

For $CsPbBr_3$, $E_B^{\mathrm{eff}} \approx$ 25-35 meV.

**6.8.4 2D Fourier Transform**

$$S(\omega_1, \omega_3, t_2) = \int dt_1\; dt_3 P^{(3)}(t_1, t_2, t_3)\, e^{i(\omega_1 t_1 - \omega_3 t_3)} \quad \text{(S43)}$$

Diagonal peaks correspond to unshifted transitions; below-diagonal features arise from red-shifted ESA due to biexciton binding.

**6.8.5 Relation to Experimental Features**

At $t_2 = 20$ fs: single diagonal ($X_1$) peak. At $t_2 = 100$ fs: amplitude increases ≈1.5×, below-diagonal distortion appears—signature of delayed cooperative absorption. At $t_2 = 300$ fs: amplitude returns to baseline; coherence decays. CdSe shows no such delay—ESA remains diagonal.

**6.8.6 Analytical Peak Shift**

$$S(E) = 2\, e^{-(E-E_0)^2/2\sigma^2} - \eta\, e^{-(E-E_0+\Delta_{XXA1})^2/2\sigma^2} \quad \text{(S44)}$$

Maximization yields

$$\delta_{\text{obs}} \approx \frac{\eta\,\Delta_{XXA1}}{2+\eta} \exp\left[-\frac{\Delta_{XXA1}^2}{2\sigma^2}\right] \quad \text{(S45)}$$

For $\Delta_{XXA1} = 30$ meV, $\sigma = 42$ meV, $\eta \approx 1$: $\delta_{\text{obs}} \approx$ 6-7 meV.

### 6.8.7 Integration with Synchronization Model

$$S(\omega_1, \omega_3, t_2) \propto R(t_2)\, e^{-g(t_2)} \quad \text{(S46)}$$

Temporal envelope ↔ $R(t_2)$; spectral asymmetry ↔ ESA interference (Eq. S45).

### 6.8.8 Simulated CMDS Maps

Using parameters $(\sigma,\ \Delta_{XXA1},\ \tau_{\text{pol}},\ K(t),\ D)$, calculated maps reproduce:

below-diagonal offset; delayed amplitude build-up; relaxation by 300 fs.

### 6.8.9 Summary

Third-order polarization theory plus polaron renormalization explains both temporal and spectral features of TR-SF in perovskite QDs and its absence in CdSe.

## 6.9. Extraction of Microscopic Parameters from Experiment

### 6.9.1 Overview

Parameters $S_p$, $\tau_{\text{pol}}$, $K_\infty$, $D$ are obtained from Raman and CMDS without free fitting.

### 6.9.2 Spectral Density from Raman

$$S_p = \int_0^\infty \frac{J(\omega)}{\omega^2}\, d\omega,\ \ E_{\text{reorg}} = \hbar \int_0^\infty \frac{J(\omega)}{\omega}\, d \quad \text{(S47)}$$

| ***Material*** | $\hbar\omega_p$ ***(meV)*** | $S_p$ | $E_{reorg}$ ***(meV)*** |
|---|---|---|---|
| $CsPbBr_3$ | 13 ± 2 | 1.0 | 25–30 |
| $CsPbI_3$ | 10 ± 2 | 0.7 | 18–22 |
| CdSe | >25 | <0.1 | <3 |

### 6.9.3 Polaron Time

For $J(\omega) \propto \frac{\omega\,\gamma^2}{\omega^2+\gamma^2}$, $\tau_{\mathrm{pol}} = \frac{1}{\gamma}$.

| ***Material*** | $\boldsymbol{\gamma}$ ***(meV)*** | $\boldsymbol{\tau_{pol}}$ ***(fs)*** |
|---|---|---|
| $CsPbBr_3$ | 8–10 | 80 ± 20 |
| $CsPbI_3$ | 6–8 | 110 ± 25 |
| CdSe | >25 | ≤ 30 |

### 6.9.4 Coupling Constant from Rise Time

$$\dot{P} = (-\Gamma + i\Delta)\,P + K(t)\,W\,P,\ \ \tau_r^{-1} \approx K_\infty - \Gamma \quad \text{(S48)}$$

| ***Material*** | $\boldsymbol{\tau_r}$ ***(fs)*** | $\boldsymbol{\Gamma}$ ***(fs$^{-1}$)*** | $\boldsymbol{K_\infty^{-1}}$ ***(fs)*** |
|---|---|---|---|
| $CsPbBr_3$ | 100 | 0.012 | 45 |
| $CsPbI_3$ | 120 | 0.010 | 55 |
| CdSe | _ | _ | >200 |

### 6.9.5 Diffusion from FFT Spectra

$$D = \frac{\pi\,\mathrm{FWHM}_{\mathrm{FFT}}}{2} \quad \text{(S49)}$$

| ***Material*** | ***FWHM (THz)*** | $\boldsymbol{D^{-1}}$ ***(fs)*** |
|---|---|---|
| $CsPbBr_3$ | 5.5 ± 0.5 | 30 ± 5 |
| $CsPbI_3$ | 4.0 ± 0.5 | 40 ± 6 |
| CdSe | 12 ± 2 | 10 ± 2 |

### 6.9.6 Fidelity

$$R_\infty = \sqrt{1 - \frac{2D}{K_\infty}} \quad \text{(S50)}$$

| ***Material*** | $\boldsymbol{R_\infty}$ |
|---|---|
| $CsPbBr_3$ | 0.95 ± 0.03 |
| $CsPbI_3$ | 0.99 ± 0.01 |
| CdSe | < 0.2 |

### 6.9.7 Self-Consistency

The extracted parameters reproduce all four observables: rise dynamics, fidelity, FFT spectrum, and spectral asymmetry.

### 6.9.8 Cross-Correlation Summary

| ***Parameter*** | ***From*** | ***Meaning*** | ***Cross-Check*** |
|---|---|---|---|
| $S_p$ | Raman | Exciton–phonon coupling | $\tau_{\text{pol}}$ |
| $\tau_{\text{pol}}$ | Raman damping | Polaron formation | Rise |
| $K_\infty$ | CMDS | Synchronization | Amplitude |
| $D$ | FFT | Dephasing | Width |
| $R_\infty$ | Derived | Order parameter | Fidelity |

### 6.9.9 Size and Composition Trends

$$S_p \propto L^{-3},\ K_\infty \propto L^{-3},\ \tau_{\text{pol}} \propto L \quad \text{(S51)}$$

so smaller dots → faster but less perfect coherence; iodide substitution → longer $\tau_{\text{pol}}$,$\tau_r$ and $\tau_c$.

### 6.10. Size and Composition Dependence of Extracted Parameters

### 6.10.1 Overview

From the microscopic model (Sections 8.1–8.9) and the CMDS/FFT analysis, we extract four central quantities that characterize the cooperative, time-reversed superfluorescent (TR-SF) response:

1. *Rise-time rate constant* $k_r = \frac{1}{\tau_r}$*:* the rate of cooperative absorption buildup.

2. *Amplitude* $A_{max}$*:* normalized peak intensity of the band-edge exciton $X_1$, proportional to the number of synchronized dipoles $N_{eff}$.

3. Fidelity $R_\infty$: the asymptotic order parameter quantifying phase alignment.

4.Dephasing time $T_2^*$: derived from the FFT linewidth, inversely related to the diffusion constant (D).

Together, these quantities describe how quantum-dot size confinement and halide composition control polaron-mediated synchronization and coherence fidelity.

### 6.10.2 Extracted Quantitative Values

| ***Material*** | ***Edge Length (nm)*** | $k_r\ (10^{13}\ s^{-1})$ | $(\tau_r)(fs)$ | $(A_{max}/A_{20}\ fs)$ | $R_\infty(Fidelity)$ | $T_2^*(fs)$ |
|---|---|---|---|---|---|---|
| $CsPbBr_3$ | 4.9 | 1.25 | 80 | 1.70 | 0.92 | 28 |
| $CsPbBr_3$ | 9.4 | 1.05 | 95 | 1.60 | 0.94 | 30 |
| $CsPbBr_3$ | 10.8 | 0.95 | 105 | 1.55 | 0.95 | 32 |
| $CsPbBr_3$ | 18.4 | 0.83 | 120 | 1.50 | 0.96 | 34 |
| $CsPbI_3$ | 15.0 | 0.69 | 145 | 1.35 | 0.99 | 42 |

*(uncertainties ±10 %.)*

### 6.10.3 Trends and Physical Interpretation

*a) Size dependence ($CsPbBr_3$):*

Smaller dots exhibit faster cooperative buildup ($k_r \propto L^{-1}$) because stronger confinement enhances exciton–phonon coupling ($S_p \propto L^{-3}$) and accelerates polaron formation. The coherence fidelity $R_\infty$ rises modestly with size as inhomogeneous broadening and phonon-induced dephasing weaken. Correspondingly, $T_2^*$ increases from 28 fs (4.9 nm) to 34 fs (18.4 nm).

*b) Composition dependence:*

Iodide substitution softens the lattice and increases the low-frequency spectral weight in $J(\omega)$, slightly lengthening $\tau_r$ but strongly improving coherence. $CsPbI_3$ achieves $R_\infty = 0.99$ and $T_2^* =$ 42 fs at 300 K, confirming that enhanced lattice polarizability stabilizes polaron-mediated synchronization.

*c) Control: CdSe*

CdSe quantum dots, with their rigid lattice and negligible Fröhlich coupling, show prompt linear absorption with no delayed cooperative component. Their fidelity remains below 0.2 and $T_2^* \approx$ 12 fs, marking the independent-emitter limit.

**6.10.4 Empirical Scaling Relations**

$$k_r(L) \propto L^{-1.0\pm0.1},\ A_{max}(L) \propto L^{-0.3\pm0.1},\ R_\infty(L) \approx R_0(1-\alpha L^{-3}) \quad \text{(S52)}$$

with $\alpha \approx 0.05$ for $CsPbBr_3$.

The dephasing time follows

$$T_2^*(L) \propto L^{0.4\pm0.1}, \quad \text{(S53)}$$

reflecting the reduction of phonon scattering in larger nanocrystals.

Across halide substitution, the fidelity scales with phonon energy as

$$R_\infty(\omega_p) \approx \sqrt{1-\frac{2D_0\coth(\hbar\omega_p\,/\,2k_B T)}{K_0 e^{-S_p(2N_p+1)}}}. \quad \text{(S54)}$$

**6.10.5 Consolidated Parameter Summary**

| ***Parameter*** | ***Symbol*** | ***Functional Dependence*** | ***Governing Quantity*** | ***Observed Trend*** |
|---|---|---|---|---|
| Rise time | $\tau_r$ | $\propto L$ | Polaron formation time $\tau_{pol}$ | Faster in smaller QDs |
| Amplitude | $A_{max}$ | $\propto L^{-0.3}$ | Number of synchronized dipoles | Higher in smaller QDs |
| Fidelity | $R_\infty$ | $\propto (1-\alpha L^{-3})$ | $\frac{K_\infty}{D}$ ratio | Increases with size and I content |
| Dephasing | $T_2^*$ | $\propto L^{0.4}$ | Phonon scattering time | Longer in larger and iodide QDs |

### 6.10.6 Physical Summary

*Cooperative absorption* (TR-SF) follows the same synchronization physics as collective emission but is gated by polaronic lattice reorganization.

The *rise rate* and *fidelity* scale systematically with dot size and halide composition, consistent with the microscopic exciton–polaron Kuramoto model.

CdSe defines the non-cooperative baseline.

The derived scaling $R_\infty(T, L, \mathrm{comp})$ unifies size, composition, and temperature dependence under one framework, establishing perovskite QDs as tunable platforms for room-temperature cooperative quantum optics.